\documentclass[twocolumn,american,nofootinbib]{revtex4}
\usepackage{ae}
\usepackage{aecompl}
\usepackage[T1]{fontenc}
\usepackage[latin1]{inputenc}
\usepackage{amsmath}
\usepackage{graphicx}
\usepackage{amssymb}

\makeatletter
\usepackage{psfrag}

\usepackage{babel}
\makeatother
\begin{document}

\title{Self-reproduction in $k$-inflation}

\author{Ferdinand Helmer and Sergei Winitzki}

\address{Arnold Sommerfeld Center, Department of Physics, Ludwig-Maximilians
University, Theresienstr.~37, 80333 Munich, Germany}

\begin{abstract}
We study cosmological self-reproduction in models of inflation driven
by a scalar field $\phi$ with a noncanonical kinetic term ($k$-inflation).
We develop a general criterion for the existence of attractors and
establish conditions selecting a class of $k$-inflation models that
admit a unique attractor solution. We then consider quantum fluctuations
on the attractor background. We show that the correlation length of
the fluctuations is of order $c_{s}H^{-1}$, where $c_{s}$ is the
speed of sound. By computing the magnitude of field fluctuations,
we determine the coefficients of Fokker-Planck equations describing
the probability distribution of the spatially averaged field $\phi$.
The field fluctuations are generally large in the inflationary attractor
regime; hence, eternal self-reproduction is a generic feature of $k$-inflation.
This is established more formally by demonstrating the existence of
stationary solutions of the relevant FP equations. We also show that
there exists a (model-dependent) range $\phi_{R}<\phi<\phi_{\max}$
within which large fluctuations are likely to drive the field towards
the upper boundary $\phi=\phi_{\max}$, where the semiclassical consideration
breaks down. An exit from inflation into reheating without reaching
$\phi_{\max}$ will occur almost surely (with probability 1) only
if the initial value of $\phi$ is below $\phi_{R}$. In this way,
strong self-reproduction effects constrain models of $k$-inflation.
\end{abstract}
\maketitle

\section{Introduction}

The paradigm of $k$-inflation~\cite{Armendariz-Picon:1999rj} assumes
that the energy density in the early universe is dominated by a scalar
field $\phi$ with an effective Lagrangian $p(X,\phi)$ consisting
solely of a (noncanonical) kinetic term, \begin{equation}
p(X,\phi)=K(\phi)L(X),\quad X\equiv\frac{1}{2}\partial_{\mu}\phi\partial^{\mu}\phi,\label{eq:Kessence L}\end{equation}
where $K(\phi)$ and $L(X)$ are functions determined by the underlying
fundamental theory. The effective Lagrangian~(\ref{eq:Kessence L})
is assumed to be valid in a sufficiently wide range of values of $\phi$
and $\partial_{\mu}\phi$, as long as the energy density of the field
$\phi$ is below a suitable scale (e.g.~the Planck scale). Unlike
the {}``traditional'' models of inflation, where the dominant energy
density is due to a potential $V(\phi)$, an accelerated expansion
in $k$-inflation is driven by the kinetic energy of the scalar field.%
\footnote{An effective scalar field $\phi$ with noncanonical kinetic terms
is also used in models of $k$-\emph{essence}~(\cite{Armendariz-Picon:2000dh,Armendariz-Picon:2000ah},
and more recently e.g.~\cite{Silverstein:2003hf,Alishahiha:2004eh},
where the Lagrangian $p(X,\phi)$ is arranged to track the equation
of state of other dominant matter components and to become dominant
only at late times. In the present paper, we do not consider such
models of $k$-essence but assume that the energy density of the field
$\phi$ is dominant during inflation. %
} The form~(\ref{eq:Kessence L}) of the Lagrangian is sufficiently
general to cover many interesting cases. For instance, Lagrangians
of the form\begin{equation}
p(X,\phi)=K_{1}(\phi)X^{n_{1}}+K_{2}(\phi)X^{n_{2}},\end{equation}
where $n_{1},n_{2}$ are fixed numbers, are reduced to the Lagrangian~(\ref{eq:Kessence L})
after a suitable redefinition of the field $\phi$. 

The assumed initial conditions for $k$-inflation consist of the field
$\phi$ having small spatial gradients within a sufficiently large
initial region, and values of $\phi$ and the time derivative $\dot{\phi}$
within suitable (and wide) ranges. These initial conditions guarantee
that the energy density due to the field $\phi$ is positive and dominates
other forms of energy in the universe. The evolving homogeneous field
$\phi(t)$ rapidly approaches an attractor solution, $\phi_{*}(t)$,
which drives inflation. The presence of the attractor makes a model
largely insensitive to initial conditions, since the attractor solution
is unique and is approached exponentially quickly. In the present
paper we develop a general definition of attractor solutions and demonstrate
their existence in a wide range of models of $k$-inflation.

The back-reaction of quantum fluctuations modifies this classical
picture of cosmological evolution, adding random {}``jumps'' to
the classical {}``drift'' along the attractor line. In generic models
of potential-driven inflation, quantum fluctuations give rise to the
phenomenon of eternal self-reproduction~\cite{Vilenkin:1983xq,Starobinsky:1986fx,Linde:1986fd,Goncharov:1987ir,Linde:1993xx}.
As soon as suitable initial conditions occur in one domain, ultimately
an infinite volume of space will undergo a long period of inflation
followed by reheating. In this way, the eternal character of the self-reproduction
process significantly alleviates the problem of initial conditions
for inflation.

The main focus of this paper is to investigate self-reproduction in
generic models of $k$-inflation with Lagrangians of the form~(\ref{eq:Kessence L}).
We find that quantum {}``jumps'' generically dominate over the {}``drift,''
and that self-reproduction is generically present. Self-reproduction
in inflation can be studied using the stochastic ({}``diffusion'')
approach based on a Fokker-Planck (FP) equation (see Ref.~\cite{Linde:1993xx}
for a review). Although the FP approach depends on the choice of the
spacelike foliation of spacetime ({}``gauge''), one can use the
FP equations to compute a number of gauge-independent characteristics,
such as the fractal dimension of the inflating domain~\cite{Aryal:1987vn,Winitzki:2001np}
and the probability of reaching the end-of-inflation point $\phi=\phi_{E}$
of the configuration space. In particular, the presence of eternal
self-reproduction is a gauge-invariant statement~\cite{Winitzki:2001np,Winitzki:2005ya}.
We show that the fractal dimension of the inflating domain is generically
close to 3, confirming a strongly self-reproducing behavior in a wide
class of models of $k$-inflation.

We also investigate the {}``stationarity'' of the distribution of
$\phi$, in the sense of Refs.~\cite{Linde:1993xx,Garcia-Bellido:1993wn,Garcia-Bellido:1994ci}.
There exists a (model-dependent) boundary value $\phi_{\max}$ such
that the stochastic formalism breaks down for $\phi>\phi_{\max}$.
Imposing a boundary condition at $\phi=\phi_{\max}$, we find that
the stationary volume-weighted distribution of the field $\phi$ is
concentrated near $\phi_{\max}$. In the terminology of Ref.~\cite{Garcia-Bellido:1993wn},
this constitutes a {}``runaway diffusion'' behavior, meaning that
the volume-averaged value of the field is driven to the imposed upper
limit. Finally, we follow the evolution of the field $\phi$ along
a single comoving worldline, starting at a generic value $\phi_{0}\lesssim\phi_{\max}$.
We find that the field will either reach the boundary $\phi=\phi_{\max}$
or, with roughly equal probability, exit the inflationary regime through
the end-of-inflation (reheating) point $\phi=\phi_{E}$. An exit into
reheating becomes overwhelmingly probable only if $\phi_{0}<\phi_{R}$,
where $\phi_{R}\ll\phi_{\max}$ is a model-dependent threshold value.
Therefore, the semiclassical description of the inflationary evolution
along a single comoving worldline will remain valid only if the initial
value $\phi_{0}$ is chosen within the interval $\phi_{E}<\phi_{0}<\phi_{R}$.
It is interesting to note that a similar conclusion was reached in
a study of stochastic back-reaction effects in models of quintessence~\cite{Martin:2004ba},
where the initial value of the inflaton and the quintessence was found
to be constrained to certain intervals. We compare this situation
with the behavior of models of potential-driven inflation, where such
a threshold $\phi_{R}\ll\phi_{\max}$ is absent and the exit into
reheating has probability $1$ for all $\phi_{0}\lesssim\phi_{\max}$,
where $\phi_{\max}$ is the Planck boundary. 

Thus, the present investigation confirms that generic models of $k$-inflation
exhibit standard features of eternal inflation, and shows that strong
effects of self-reproduction constrain the choice of initial conditions.

\section{Main results\label{sec:Main-results}}

In this section we present the main line of arguments and the principal
results of this paper. Full details of the relevant derivations are
delegated to sections~\ref{sec:Attractor-solutions} to \ref{sec:Analysis-of-Fokker-Planck}.

In the cosmological context, we consider a spatially homogeneous field
$\phi=\phi(t)$ and a flat FRW metric,\begin{equation}
g_{\mu\nu}dx^{\mu}dx^{\nu}=dt^{2}-a^{2}(t)d\mathbf{x}^{2},\end{equation}
where $a(t)$ is the scale factor. We also assume that the energy
density of the field $\phi$,\begin{equation}
\varepsilon(X,\phi)\equiv2Xp_{,X}-p,\end{equation}
 dominates all the other forms of energy in the universe. (Partial
derivatives are denoted by a comma, so $p_{,X}\equiv\partial p/\partial X$.)
Then, as is well known, the Einstein equations yield\begin{equation}
\frac{\dot{a}}{a}\equiv H=\kappa\sqrt{\varepsilon},\quad\kappa^{2}\equiv\frac{8\pi G}{3}=\frac{8\pi}{3M_{Pl}^{2}},\end{equation}
while the equation of motion for the field $\phi$,\begin{equation}
\frac{d}{dt}\left(p_{,v}\right)+3Hp_{,v}=p_{,\phi},\quad v\equiv\dot{\phi}=\sqrt{2X},\label{eq:initial EOM}\end{equation}
 can be rewritten as \cite{Armendariz-Picon:1999rj}\begin{equation}
\dot{\varepsilon}=-3H\left(\varepsilon+p\right).\label{eq:epsilon EOM}\end{equation}

In Sections \ref{sub:Definition-of-attractor}-\ref{sub:Summary-of-assumptions}
we determine the conditions for the existence of attractor solutions
in general models of $k$-inflation. In a typical scenario, the field
starts at a large initial value $\phi=\phi_{0}$ with $\dot{\phi}<0$
and gradually approaches a smaller value $\phi=\phi_{E}$ where inflation
ends. If an attractor exists, the trajectory in the phase space quickly
approaches the attractor curve; the equation of motion for $\phi(t)$
is effectively reduced to a first-order equation. A slow-roll parameter
can be introduced to quantify the deviation of the spacetime from
exact de Sitter. In Sec.~\ref{sub:Existence-of-attractors} we show
that a cosmological model with a Lagrangian of the form~(\ref{eq:Kessence L})
admits an inflationary attractor solution under the following (sufficient)
conditions:

(i) The function $K(\phi)$ remains positive at $\phi\rightarrow\infty$
and is such that $\int^{\infty}\sqrt{K(\phi)}d\phi=\infty$ (the integral
diverges) and $\lim_{\phi\rightarrow\infty}(K^{-3/2}K^{\prime})=0$.
For instance, functions $K(\phi)\sim\phi^{s}$ with $s>-2$ satisfy
these two conditions.%
\footnote{Technically, the two conditions are independent, as exemplified by
the functions $K(\phi)=\phi^{-1}$ and $K(\phi)=\left(\phi\ln\phi\right)^{-2}$.
Note that the borderline case $K(\phi)=\phi^{-2}$ must be considered
separately (the attractor has the exact form $\dot{\phi}=\textrm{const}$).
Below we shall only consider $K(\phi)\sim\phi^{s}$ with $s\geq0$.%
}

(ii) The equation $L'(X)=0$ has a unique root $X_{0}>0$ such that
the energy density, $\varepsilon(X_{0},\phi)=-K(\phi)L(X_{0})$, is
positive and bounded away from zero for large $\phi$. (This condition
excludes $K(\phi)\sim\phi^{s}$ with $s<0$.) The attractor solution
is $X=X_{*}(\phi)$, where the function $X_{*}(\phi)$ tends to $X_{0}$
for large $\phi$.

(iii) The speed of sound of field perturbations is real-valued, $c_{s}^{2}>0$,
for $X=X_{0}$ and large $\phi$ (i.e.~perturbations on the attractor
background do not cause instabilities). It will be shown that $c_{s}^{2}>0$
if $L^{\prime\prime}(X_{0})K^{\prime}(\phi)>0$ for large $\phi$.
The speed of sound $c_{s}$ (which is small in typical models of $k$-inflation)
plays the role of a slow-roll parameter.

The attractor solution will be $\dot{\phi}=v_{*}(\phi)$; as long
as the system is on the attractor, all the cosmological functions
can be expressed as functions of $\phi$. We obtain a complete asymptotic
expansion of $v_{*}(\phi)$ at large $\phi$ (Sec.~\ref{sub:Asymptotics-in-the-general})
and a representation in terms of an integral equation~(Sec.~\ref{sub:Existence-of-attractors-general}).
We show that the slow-roll condition generally holds (Sec.~\ref{sub:Slow-roll-condition}).
Explicit asymptotics in the leading slow-roll approximation are given
in Sec.~\ref{sub:Asymptotics-of-attractors-in-k}.

In the stochastic approach to self-reproduction, one is interested
in averaged values of the field $\phi$ on sufficiently large distance
scales, on which the fluctuations freeze and become uncorrelated.
We show in Sec.~\ref{sub:Scales-of-averaging} that the relevant
scale $L$ is not the Hubble horizon $H^{-1}$ but the \emph{sound
horizon}, $L\sim c_{s}H^{-1}$, where the speed of sound is \begin{equation}
c_{s}(X,\phi)=\sqrt{\frac{p_{,X}}{\varepsilon_{,X}}}.\label{eq:cs def}\end{equation}
 Let us denote the averaging on these scales by an overbar. The dynamics
of the averaged field $\bar{\phi}$ is a superposition of the deterministic
motion and quantum {}``jumps'' due to the backreaction of quantum
fluctuations exiting the (sound) horizon. Since the attractor solution
is approached exponentially quickly by other solutions, any deviations
from the attractor are quickly suppressed. So we may assume that at
every place the cosmological quantities such as $H$, $c_{s}$, are
always given by the functions of $\phi$ derived from the attractor
solution (Sec.~\ref{sub:Asymptotics-of-attractors-in-k}).

The backreaction of quantum fluctuations on the averaged field $\bar{\phi}$
may lead to a qualitative change in the evolution of the field, compared
with the deterministic trajectory $\dot{\phi}=v_{*}(\phi)$. To describe
the dynamics of the averaged field during inflation, we use the stochastic
or {}``diffusion'' approach (see Ref.~\cite{Linde:1993xx} for
a review). The distribution $P_{c}(\phi,t)$ of values of $\bar{\phi}$
per \emph{coordinate} (or comoving) volume satisfies the Fokker-Planck
(FP) equation,\begin{equation}
\frac{\partial P_{c}}{\partial t}=\frac{\partial^{2}}{\partial\phi^{2}}\left(DP_{c}\right)-\frac{\partial}{\partial\phi}\left(v_{*}P_{c}\right)\equiv\hat{L}_{c}P,\label{eq:FP equ c}\end{equation}
where $v_{*}(\phi)$ is the deterministic attractor trajectory and
$D(\phi)$ is the {}``diffusion coefficient'' describing the magnitude
of fluctuations. The computation of $D(\phi)$ is performed in Sec.~\ref{sub:Diffusion-coefficient},
and the result (in the leading slow-roll approximation) is expressed
as \begin{equation}
D(\phi)\approx\textrm{const}\frac{H(\phi)}{c_{s}^{3}(\phi)}.\end{equation}

The comoving distribution $P_{c}$ can be interpreted as the probability
density, $P_{c}(\phi,t)d\phi$, for the value of $\phi$ at time $t$
along a randomly chosen comoving worldline $\mathbf{x}=\textrm{const}$.
Also of interest is the distribution of $\bar{\phi}$ weighted by
\emph{proper} 3-volume within 3-surfaces of equal time $t$. This
volume-weighted distribution, denoted $P_{p}(\phi,t)$, satisfies
an analogous FP equation, which differs from Eq.~(\ref{eq:FP equ c})
only by the presence of the expansion term $3H(\phi)P$,\begin{equation}
\frac{\partial P_{p}}{\partial t}=\frac{\partial^{2}}{\partial\phi^{2}}\left(DP_{p}\right)-\frac{\partial}{\partial\phi}\left(v_{*}P_{p}\right)+3HP_{p}\equiv\hat{L}_{p}P.\label{eq:FP equ}\end{equation}
 These FP equations are supplemented by boundary conditions. One imposes
an {}``exit-only'' condition at the end-of-inflation point, $\phi=\phi_{E}$,
and either a reflecting or an absorbing boundary condition at a suitable
value $\phi=\phi_{\max}$ (see Sec.~\ref{sub:Eigenvalues-of-the-comoving-FP}
for explicit details). One then expects that solutions of Eq.~(\ref{eq:FP equ c})
should decay with time, indicating that at late times a given comoving
worldline will have almost surely exited the inflationary regime.
In Sec.~\ref{sub:Eigenvalues-of-the-comoving-FP}, we show that the
differential operator $\hat{L}_{c}$ in the r.h.s.~of Eq.~(\ref{eq:FP equ c})
indeed has only negative eigenvalues, indicating a late-time behavior
$P_{c}(\phi,t)\propto e^{-\lambda t}$ with $\lambda>0$. On the other
hand, solutions of Eq.~(\ref{eq:FP equ}) may either decay or grow
at late times, depending on whether the largest eigenvalue $\lambda_{\max}$
of the operator $\hat{L}_{p}$ is negative or positive. A positive
value of $\lambda_{\max}$ indicates the presence of eternal self-reproduction
(eternal inflation), with a stationary distribution $P_{p}(\phi,t)\propto e^{\lambda_{\max}t}$.
It is found in Sec.~\ref{sub:Stationary-solutions-in-k} that $\lambda_{\max}>0$
in generic models of $k$-inflation, and that the distribution $P_{p}(\phi,t)$
is concentrated near the upper boundary $\phi=\phi_{\max}$. Moreover,
we show that $\lambda_{\max}\approx3H_{\max}$, where $H_{\max}=H(\phi_{\max})$
is the largest value of $H(\phi)$ within the allowed range of $\phi$.
This shows that the inflating domain grows almost at the rate of the
Hubble expansion. It is known that the inflating domain can be visualized
as a self-similar fractal set~\cite{Aryal:1987vn,Winitzki:2001np}.
The fact that $\lambda_{\max}\approx3H_{\max}$ means that the fractal
dimension of the inflating domain is close to 3, indicating a regime
of strong self-reproduction.

Finally, we calculate the probability of reaching the end-of-inflation
boundary $\phi_{E}$ if the evolution starts at an initial value $\phi=\phi_{0}$
(Sec.~\ref{sub:Exit-probability-in}). In all models of $k$-inflation,
the consistency of the diffusion approach requires one to limit the
allowed range of $\phi$ by an upper boundary $\phi_{\max}$, even
if the energy density does not reach the Planck scale (Sec.~\ref{sub:Validity-of-diffusion}).
The boundary $\phi_{\max}$ is analogous to the Planck boundary in
models of potential-driven chaotic inflation, in that the semiclassical
({}``diffusion'') approach fails for $\phi>\phi_{\max}$ because
fluctuations become too large. One expects that the probability of
exiting through the end-of-inflation point $\phi_{E}$, rather than
through the upper boundary $\phi_{\max}$, should be almost equal
to 1, indicating that almost all comoving observers will exit inflation
normally. We show that there exists a threshold value $\phi_{R}\ll\phi_{\max}$
such that the exit through $\phi_{E}$ is overwhelmingly likely for
initial values $\phi_{0}<\phi_{R}$. (Such a threshold is absent in
potential-driven inflation.) Hence, a given comoving worldline with
$\phi_{0}<\phi_{R}$ will almost surely (with probability $1$) eventually
exit inflation and enter the reheating phase. Nevertheless, the total
proper 3-volume of the inflating domains will grow exponentially with
time. Thus, we demonstrate that generic models of $k$-inflation exhibit
the standard features of eternal self-reproduction.

\section{Attractor solutions in $k$-inflation\label{sec:Attractor-solutions}}

In this section we study the cosmological dynamics of the spatially
homogeneous field $\phi(t)$ with a Lagrangian $p(\dot{\phi},\phi)$
and investigate the existence of attractor solutions in the regime
$\phi\rightarrow\infty$. We show that attractor solutions exist in
a broad class of models and determine the asymptotic forms of the
attractors. As a particular example, we select the Lagrangian\begin{equation}
p(\dot{\phi},\phi)=K(\phi)Q(\dot{\phi}).\label{eq:KEssense p}\end{equation}

Since the equation of motion~(\ref{eq:initial EOM}) does not depend
explicitly on time $t$, it is convenient to consider solutions $\phi(t)$
as curves in the phase plane $\left(\phi,v\right)$, where we denote
$v\equiv\dot{\phi}$. The equation of motion in the phase plane has
the form\begin{equation}
\frac{dv}{d\phi}=g(v,\phi),\label{eq:v g EOM}\end{equation}
where $g(v,\phi)$ is an auxiliary function expressed through the
Lagrangian $p(v,\phi)$ as\begin{align}
g(v,\phi) & =-\frac{1}{vp_{,vv}}\left(3\kappa\sqrt{\varepsilon(v,\phi)}p_{,v}+vp_{,v\phi}-p_{,\phi}\right),\label{eq:g def}\\
\varepsilon(v,\phi) & \equiv vp_{,v}-p.\end{align}
 In the particular case of the Lagrangian~(\ref{eq:KEssense p}),
the function $g(v,\phi)$ is given by\begin{align}
g(v,\phi) & =-3\kappa\frac{Q^{\prime}(v)\sqrt{\tilde{\varepsilon}(v)}}{vQ^{\prime\prime}(v)}\sqrt{K(\phi)}-\frac{\tilde{\varepsilon}(v)}{vQ^{\prime\prime}(v)}\frac{K^{\prime}(\phi)}{K(\phi)},\label{eq:g for KEssence}\\
\tilde{\varepsilon}(v) & \equiv\frac{\varepsilon(v,\phi)}{K(\phi)}=vQ^{\prime}(v)-Q.\end{align}
Here and below we assume that $K(\phi)>0$ when $\phi\rightarrow\infty$.
We can always change the sign of $Q(\dot{\phi})$ if this is not the
case. Note that $K(\phi)$ may have a root at a finite $\phi$, e.g.~at
the endpoint of inflation, but we only consider the inflationary regime
where $\phi$ is sufficiently large and $K(\phi)$ remains positive.

\subsection{Definition of attractors\label{sub:Definition-of-attractor}}

\begin{figure}
\begin{center}\psfrag{f}{$\phi$}\psfrag{v}{$v$}\psfrag{v0}{$v=v_0$}\includegraphics[%
  width=3.3in]{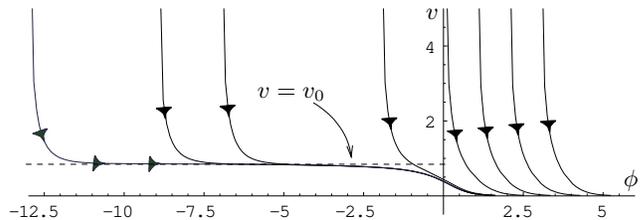}\end{center}

\caption{A numerically obtained phase plot for the Lagrangian $p(X,\phi)=(-X+X^{2})\arctan\phi$.
Trajectories starting at large negative $\phi$ with large positive
$v\equiv\dot{\phi}$ approach an attractor solution. The attractor
trajectory approximately coincides with the line $v=v_{0}$ for sufficiently
large $\left|\phi\right|$.\label{cap:Phase-plot-1}}
\end{figure}

Let us now motivate the definition of an attractor by analyzing the
typical behavior of solutions in terms of the function $v(\phi)$.
A sample set of trajectories in the phase plane $\left(\phi,v\right)$
is shown in Fig.~\ref{cap:Phase-plot-1}. Trajectories starting at
large $\phi$ and $v<0$ will quickly approach an almost horizontal
line, $v\approx v_{0}=\textrm{const}$, and then proceed along that
line more slowly towards $\phi=0$. It is intuitively clear from the
figure that the line $v\approx v_{0}$ (rather than any of the neighbor
trajectories) should be considered the attractor solution. To make
this statement precise, we need a formal criterion that would distinguish
the attractor line from nearby solutions.

We begin by presenting a heuristic motivation for this criterion (which
will be Eq.~(\ref{eq:attractor def}) below). In Fig.~\ref{cap:Phase-plot-1},
the attractor solution stays approximately constant as $\phi\rightarrow-\infty$,
while a generic neighbor trajectory $v_{0}+\delta v(\phi)$ quickly
moves away from the attractor. Let us therefore examine the growth
of $\delta v(\phi)$ with $\phi$ as $\phi\rightarrow-\infty$. If
$v\approx v_{0}$ is a solution of Eq.~(\ref{eq:v g EOM}), then
$g(v_{0},\phi)\approx0$ and $g(v_{0}+\delta v,\phi)\approx g_{,v}(v_{0},\phi)\delta v$.
Since $\frac{d}{d\phi}\delta v\approx g_{,v}\delta v$, an initially
small deviation $\delta v(\phi)$ grows exponentially with $\phi$
if $g_{,v}(v_{0},\phi)\neq0$. Assuming a power-law dependence $p(v,\phi)\sim v^{n}$
for large $v$, one finds from Eq.~(\ref{eq:g def}) that the function
$g(v,\phi)$ will grow at least linearly with $v$ at fixed $\phi$;
the growth of $g(v,\phi)$ with $v$ will be even faster if $p(v,\phi)$
is exponential in $v$. Hence, the growth of a non-attractor solution
for large $\phi$ is at least exponential, both for small and for
large deviations from the attractor.

For convenience, below we always consider the quadrant $\phi>0$ and
$v<0$ in the phase plane, since the opposite choice $\phi\rightarrow-\infty$
and $v>0$ is treated completely analogously. In the general case,
we expect that the attractor solution $v_{*}(\phi)$ will have a relatively
slow behavior at $\phi\rightarrow\infty$, compared with an exponential
behavior of nearby non-attractor solutions. Thus, a suitable condition
for selecting the attractor solution is that $v_{*}(\phi)$ should
grow slower than exponentially as $\phi\rightarrow\infty$, while
nearby solutions grow exponentially. A formal way to express this
condition is\begin{equation}
\lim_{\phi\rightarrow\infty}\frac{d}{d\phi}\ln v_{*}(\phi)=0.\label{eq:attractor def}\end{equation}
We shall use Eqs.~(\ref{eq:v g EOM}) and (\ref{eq:attractor def})
as the \emph{definition} of the attractor solution $v_{*}(\phi)$,
together with the assumption that nearby trajectories do \emph{not}
satisfy Eq.~(\ref{eq:attractor def}). We stress that the attractor
solution is singled out by an asymptotic condition at $\phi\rightarrow\infty$.
At a finite value of $\phi$, all the trajectories approach each other
and no single solution appears to be special.

Let us remark that the definition~(\ref{eq:attractor def}) may be
unduly restrictive. In particular, it does not allow attractors that
are approached more slowly than exponentially. In such cases, the
definition~(\ref{eq:attractor def}) may still be used to select
the attractor solution after a suitable change of variable $\phi\rightarrow\tilde{\phi}$.
Moreover, it may happen that either every solution or no solution
of Eq.~(\ref{eq:v g EOM}) satisfies Eq.~(\ref{eq:attractor def}).
In such cases, a more detailed analysis is required to investigate
the existence of attractors. Nevertheless, the condition~(\ref{eq:attractor def})
is adequate for the analysis of $k$-inflation.

\subsection{Asymptotics in the general case\label{sub:Asymptotics-in-the-general}}

The next task is to determine the possible asymptotic behavior of
the attractor $v_{*}(\phi)$, given the equation of motion~(\ref{eq:v g EOM}).
Let us assume that the function $g(v,\phi)$ has an asymptotic expansion
at $\phi\rightarrow\infty$ of the form\begin{equation}
g(v,\phi)=A_{0}(v)B_{0}(\phi)+A_{1}(v)B_{1}(\phi)+...,\label{eq:asympt of g}\end{equation}
where $B_{1}(\phi)$ is subdominant to $B_{0}(\phi)$ as $\phi\rightarrow\infty$.
The asymptotic expansion of $g(v,\phi)$ at $\phi\rightarrow\infty$
will have the form~(\ref{eq:asympt of g}) if the Lagrangian $p(v,\phi)$
is polynomial in $v$. In particular, Eq.~(\ref{eq:g for KEssence})
is manifestly in this form. 

In most cases, the term $A_{0}(v)B_{0}(\phi)$ in Eq.~(\ref{eq:asympt of g})
will dominate at large $\phi$, so that an approximate \emph{general}
(i.e.~non-attractor) solution $v(\phi)$ of Eq.~(\ref{eq:v g EOM})
is found as \begin{equation}
\int_{v(\phi)}^{v_{1}}\frac{dv'}{A_{0}(v')}=\int_{\phi}^{\phi_{1}}B_{0}(\phi')d\phi',\label{eq:approx gen sol}\end{equation}
where $v(\phi_{1})=v_{1}$ is an arbitrary initial condition. The
technical assumptions for the dominance of the term $A_{0}(v)B_{0}(\phi)$
are that both $A_{0}(v)$ and $A_{1}(v)$ are bounded for finite $v$
and that $A_{1}(v)$ does not grow faster than $A_{0}(v)$ for large
$v$. It is straightforward to verify, using Eq.~(\ref{eq:g def}),
that these assumptions hold for a Lagrangian $p(\phi,v)$ which is
polynomial in $v$. 

To study the behavior of the general solution~(\ref{eq:approx gen sol})
for large $\phi$, we need to consider two mutually exclusive cases:
either $\int^{\infty}B_{0}(\phi)d\phi$ converges at $\phi\rightarrow\infty$,
or it diverges. If this integral converges, there exist solutions
$v(\phi)$ that approach any given value $v_{1}$ as $\phi\rightarrow\infty$,
namely,\begin{equation}
\int_{v(\phi)}^{v_{1}}\frac{dv'}{A_{0}(v')}=\int_{\phi}^{\infty}B_{0}(\phi')d\phi'.\end{equation}
Since $v_{1}$ can be chosen continuously, there are infinitely many
solutions $v(t)$ that tend to a constant limit at $\phi\rightarrow\infty$.
It follows that there is no unique attractor solution in this case.
In the second case, $\int^{\infty}B_{0}(\phi)d\phi=\infty$, there
may exist solutions $v(\phi)$ approaching a root of $A_{0}(v)$ as
$\phi\rightarrow\infty$. To simplify the analysis, let us change
the variables, \begin{equation}
\phi\rightarrow\tilde{\phi}(\phi)\equiv\int^{\phi}B_{0}(\phi)d\phi,\end{equation}
 noting that $\tilde{\phi}\rightarrow\infty$ together with $\phi\rightarrow\infty$.
We can now rewrite the equation of motion through the new variable
$\tilde{\phi}$ and look for attractor solutions at $\tilde{\phi}\rightarrow\infty$
using the same criteria as before. The asymptotic form of $\tilde{g}(v,\tilde{\phi})\equiv g(v,\phi)/B_{0}(\phi)$
will be\begin{equation}
\tilde{g}(v,\tilde{\phi})=A_{0}(v)+A_{1}(v)\left.\frac{B_{1}(\phi)}{B_{0}(\phi)}\right|_{\phi\rightarrow\tilde{\phi}}+...\end{equation}
Essentially, the replacement $\phi\rightarrow\tilde{\phi}$ is equivalent
to the assumption that $B_{0}(\phi)\equiv1$ in Eq.~(\ref{eq:asympt of g}).
This is what we shall assume in this and the following sections. 

If $v_{*}(\phi)$ is an attractor, Eq.~(\ref{eq:attractor def})
with $B_{0}(\phi)\equiv1$ gives\begin{equation}
0=\lim_{\phi\rightarrow\infty}\frac{g(v_{*}(\phi),\phi)}{v_{*}(\phi)}=\lim_{\phi\rightarrow\infty}\frac{A_{0}(v_{*}(\phi))}{v_{*}(\phi)},\end{equation}
and hence $v_{*}(\phi)$ approaches a root of $A_{0}(v)/v$ as $\phi\rightarrow\infty$.
Thus, roots of the function $A_{0}(v)/v$ correspond to solutions
that are approximately constant for large $\phi$. Our previous analysis
shows that such solutions, $v(\phi)\approx v_{0}$, will be attractors
if $A_{0}^{\prime}(v_{0})>0$ and repulsors if $A_{0}^{\prime}(v_{0})<0$.
Therefore, we focus attention on the case when the function $A_{0}(v)$
has a finite root $v_{0}<0$ such that $A_{0}^{\prime}(v_{0})>0$,
and we look for an attractor solution $v_{*}(\phi)$ that approaches
$v=v_{0}$ when $\phi\rightarrow\infty$.%
\footnote{Thus, we omit the consideration of cases where the existence of attractors
is less straightforward to establish. For instance, the function $A_{0}(v)/v$
may have no finite roots, may approach zero as $v\rightarrow\infty$,
may have a root $v_{0}=0$, or a higher-order (multiple) root. Also,
an attractor solution may exist not due to a root of $A_{0}(v)$ but
due to a cancellation among several terms in the expansion~(\ref{eq:asympt of g}).
Finally, an attractor solution may grow without bound, $v_{*}(\phi)\rightarrow\infty$
as $\phi\rightarrow\infty$, instead of approaching a constant. In
every such case, a more detailed analysis is needed to investigate
the existence of attractors, possibly involving a further change of
variable, such as $v(\phi)\equiv\tilde{v}(\phi)F(\phi)$, where $F(\phi)$
is a function with a suitable asymptotic behavior at $\phi\rightarrow\infty$.
See, for instance, Appendix~\ref{sec:Attractor-solutions-in} where
we use this technique to establish the existence of unique attractors
in models of potential-driven inflation.%
} Additionally, a physically meaningful inflationary attractor solution
$v_{*}(\phi)$ must have a positive energy density, $\varepsilon(v_{*}(\phi),\phi)>0$.
Thus the root $v_{0}$ must also satisfy the condition\begin{equation}
\lim_{\phi\rightarrow\infty}\varepsilon(v_{0},\phi)>0.\end{equation}

Since $v_{*}(\phi)\approx\textrm{const}$ for large $\phi$, an approximation
to the attractor solution can be found heuristically by neglecting
$v'$ in Eq.~(\ref{eq:v g EOM}). Clearly, this is equivalent to
the frequently used \emph{slow-roll approximation} where one neglects
$\ddot{\phi}$ and keeps only $\dot{\phi}$ in the equations of motion
for $\phi(t)$. Thus, the slow-roll approximation $v_{sr}(\phi)$
to the attractor solution is the root of $g(v_{sr}(\phi),\phi)=0$
which is near $v_{0}$ as $\phi\rightarrow\infty$. 

We can now obtain a complete asymptotic expansion of the attractor
solution $v_{*}(\phi)$ at $\phi\rightarrow\infty$. Since $\lim_{\phi\rightarrow\infty}g_{,v}(v,\phi)=A_{0}^{\prime}(v)\neq0$
in a neighborhood of $v_{0}$, we may solve Eq.~(\ref{eq:v g EOM})
with respect to $v$ at least in some neighborhood of $(\phi=\infty,v=v_{0})$
and obtain \begin{equation}
v(\phi)=g^{-1}(\frac{dv}{d\phi},\phi),\label{eq:inverse iteration}\end{equation}
where $g^{-1}(\cdot,\phi)$ is the inverse of $g(v,\phi)$ with respect
to $v$. (If the equation $g(v,\phi)=0$ has several roots with respect
to $v$, an appropriate choice of the branch of the inverse function
must be made to ensure that $g^{-1}(0,\phi)\rightarrow v_{0}$ as
$\phi\rightarrow\infty$). Let us apply the method of iteration to
Eq.~(\ref{eq:inverse iteration}), starting with the initial approximation
$v_{(0)}(\phi)=\textrm{const}$. The first approximation will coincide
with the slow-roll solution, \begin{equation}
v_{(1)}(\phi)=g^{-1}(0,\phi)\equiv v_{sr}(\phi).\end{equation}
We shall now show that the successive approximations $v_{(n)}(\phi)$,
$n=1,2,...$, differ from the exact attractor by terms of progressively
higher asymptotic order at $\phi\rightarrow\infty$. It will follow
that the sequence $\left\{ v_{n}(\phi)\right\} $ converges to the
attractor in the \emph{asymptotic} sense (i.e.~for $\phi\rightarrow\infty$
at fixed $n$). This asymptotic approximation is similar in spirit
to the asymptotic expansion using a hierarchy of slow-roll parameters,
which was developed in Ref.~\cite{Liddle:1994dx}. 

By construction, successive approximations $v_{(n)}(\phi)$ satisfy\begin{equation}
v_{(n+1)}(\phi)=g^{-1}\left(\frac{dv_{(n)}(\phi)}{d\phi},\phi\right).\end{equation}
 We denote by $v_{*}(\phi)$ the \emph{exact} attractor solution satisfying
Eq.~(\ref{eq:inverse iteration}). Then the deviation of $v_{(n)}$
from $v_{*}$ satisfies the recurrence relation\begin{align}
 & v_{(n+1)}-v_{*}=g^{-1}(v_{(n)}^{\prime},\phi)-g^{-1}(v_{*}^{\prime},\phi)\nonumber \\
 & =\frac{1}{g_{,v}(v_{*},\phi)}\left[v_{(n)}-v_{*}\right]_{,\phi}+O\left[\left(v_{(n)}-v_{*}\right)^{2}\right].\end{align}
Since $v_{(1)}(\phi)-v_{*}(\phi)$ decays at $\phi\rightarrow\infty$
(this is so because both functions approach the same constant $v_{0}$),
it follows that $(v_{(n)}-v_{*})$ also \emph{decays} at $\phi\rightarrow\infty$
for all $n$. Hence, the terms $(v_{(n)}-v_{*})_{,\phi}$ and $(v_{(n)}-v_{*})^{2}$
have a higher asymptotic order (i.e.~decay faster as $\phi\rightarrow\infty$)
than $(v_{(n)}-v_{*})$. Since for large $\phi$ we have $g_{,v}(v_{*})\approx A_{0}^{\prime}(v_{0})$,
which is a nonzero constant, it follows that $(v_{(n+1)}-v_{*})$
has a higher asymptotic order than $(v_{(n)}-v_{*})$ at $\phi\rightarrow\infty$.

\subsection{Existence of attractors\label{sub:Existence-of-attractors-general}}

In the previous section, we have shown that an asymptotic expansion
of the attractor solution may be obtained by iterating Eq.~(\ref{eq:inverse iteration}).
However, the successive terms of the expansion will contain derivatives
of $B_{1}(\phi)$ of successively higher orders. Since the $n$-th
derivative of an analytic function usually grows as quickly as $\sim n!$,
the asymptotic expansion could be a divergent series. Such a series
cannot be used to reconstruct the attractor solution exactly and establish
its existence. In fact, the asymptotic property of the sequence $v_{(n)}(\phi)$
was derived \emph{assuming} that the exact attractor solution $v_{*}(\phi)$
exists. 

The existence of the attractor solution can be established by producing
a convergent sequence based on an integral equation. To this end,
we rewrite the function $g(v,\phi)$ as\begin{equation}
g(v,\phi)=\left(v-v_{0}\right)\alpha+\left(v-v_{0}\right)^{2}A(v)+B(v,\phi),\end{equation}
where $\alpha\equiv A_{0}^{\prime}(v_{0})>0$ and $A(v),B(v,\phi)$
are auxiliary functions such that $A(v)$ is regular at $v=v_{0}$
and $B(v,\phi)$ decays as $\phi\rightarrow\infty$ uniformly for
all $v$ near $v_{0}$. (It is possible to express $g(v,\phi)$ in
the above form because $g_{,v}(v_{0},\phi)\rightarrow\alpha$ as $\phi\rightarrow\infty$.
The function $B(v,\phi)$ has the stated property since $A_{1}(v)$
is regular at $v_{0}$.) A trivial redefinition $\phi\rightarrow\alpha^{-1}\phi$
effectively sets $\alpha=1$, which simplifies the analysis. Then
Eq.~(\ref{eq:v g EOM}) with the boundary condition $v(\infty)=v_{0}$
is equivalent to the following integral equation,\begin{align}
v(\phi)=v_{0}-e^{\phi}\int_{\phi}^{\infty}e^{-\phi'} & \big[\left(v(\phi')-v_{0}\right)^{2}A(v(\phi'))\nonumber \\
 & +B(v(\phi'),\phi')\big]d\phi'.\label{eq:vstar integral equ}\end{align}
This equation can be iterated starting from the initial approximation
$v^{(0)}(\phi)\equiv v_{0}$, which produces successive approximations
$v^{(n)}(\phi)$, $n=1,2,...$ We shall now show that the sequence
$v^{(n)}(\phi)$ converges as $n\rightarrow\infty$ for sufficiently
large (but fixed) $\phi$. The limit of the sequence will be, evidently,
the attractor solution $v_{*}(\phi)$.

We shall prove the convergence of the sequence $\left\{ v^{(n)}(\phi)\right\} $
by showing that the difference between successive approximations,
$\left|v^{(n+1)}(\phi)-v^{(n)}(\phi)\right|$, decays with $n$ at
fixed $\phi$. Let us first estimate the difference $\Delta_{n}(\phi)\equiv v^{(n)}(\phi)-v_{0}$
at some fixed $\phi$. By construction, we have\begin{align}
 & \Delta_{n+1}(\phi)\nonumber \\
 & =-e^{\phi}\int_{\phi}^{\infty}e^{-\phi'}\left[\Delta_{n}^{2}A(v^{(n)}(\phi'))+B(v^{(n)}(\phi'),\phi')\right]d\phi'.\label{eq:Deltan integral equ}\end{align}
 To determine a bound for $\Delta_{n}$ using this equation, we need
some bounds on the functions $A(v)$ and $B(v,\phi)$ valid near $v=v_{0}$
and for large $\phi$. By assumption, $B(v,\phi)$ decays uniformly
in $\phi,$ so for any small number $\delta>0$ we can choose $\phi$
large enough, say $\phi>\phi_{0}$, so that \begin{equation}
\left.\begin{array}{r}
\left|B(v,\phi)\right|<\delta\\
\left|B_{,v}(v,\phi)\right|<\delta\end{array}\right\} \quad\textrm{for }\left|v-v_{0}\right|<\delta\;\textrm{and }\phi>\phi_{0}.\label{eq:delta phi0 cond}\end{equation}
Since $A(v)$ is regular at $v=v_{0}$, there exists a constant $C>0$
such that $\left|A(v)\right|<C$ for $\left|v-v_{0}\right|<\delta$.
Then the difference $\Delta_{n+1}$ is estimated using Eq.~(\ref{eq:Deltan integral equ})
as follows,\begin{equation}
\left|\Delta_{n+1}(\phi)\right|\leq C\left|\Delta_{n}(\phi)\right|^{2}+\delta.\end{equation}
Since the initial approximation is $v^{(0)}(\phi)\equiv v_{0}$, it
is easy to see that, with $\delta<(4C)^{-1}$, we have\begin{equation}
\left|\Delta_{n}(\phi)\right|<\frac{1-\sqrt{1-4C\delta}}{2C}<2\delta,\;\textrm{for }\forall n\textrm{ and }\phi>\phi_{0}.\end{equation}
 Let us choose $\delta<(4C)^{-1}$ and find a corresponding value
of $\phi_{0}$ such that Eq.~(\ref{eq:delta phi0 cond}) holds. Then
we compute\begin{align}
 & \left|v^{(n+1)}(\phi)-v^{(n)}(\phi)\right|\nonumber \\
 & \leq\int_{\phi}^{\infty}e^{\phi-\phi'}\left[\left|(v^{(n)}-v_{0})^{2}-(v^{(n-1)}-v_{0})^{2}\right|C\right.\nonumber \\
 & \left.\quad+\left|v^{(n)}-v^{(n-1)}\right|\delta\right]d\phi'.\end{align}
Since\begin{align}
 & \left|(v^{(n)}-v_{0})^{2}-(v^{(n-1)}-v_{0})^{2}\right|\nonumber \\
 & =\left|(v^{(n)}-v^{(n-1)})(v^{(n)}-v_{0}+v^{(n-1)}-v_{0})\right|\nonumber \\
 & <4\left|v^{(n)}-v^{(n-1)}\right|\delta,\end{align}
we have\begin{align}
 & \left|v^{(n+1)}(\phi)-v^{(n)}(\phi)\right|\nonumber \\
 & \leq5\delta\int_{\phi}^{\infty}e^{\phi-\phi'}\left|v^{(n)}(\phi')-v^{(n-1)}(\phi')\right|d\phi'.\end{align}
Now we use induction in $n$, starting with \begin{align}
\left|v^{(1)}(\phi)-v^{(0)}(\phi)\right| & \leq\int_{\phi}^{\infty}d\phi\, e^{\phi-\phi'}B(v_{0},\phi)\nonumber \\
 & <\int_{\phi}^{\infty}d\phi\, e^{\phi-\phi'}\delta=\delta,\end{align}
to establish the bound \begin{equation}
\left|v^{(n+1)}(\phi)-v^{(n)}(\phi)\right|<\left(5\delta\right)^{n}\delta,\;\textrm{for }\forall n\textrm{ and }\phi>\phi_{0}.\end{equation}
This bound tends to zero for $n\rightarrow\infty$ as long as $\delta<\frac{1}{5}$.
If necessary, we can diminish $\delta$ so that $\delta<\frac{1}{5}$
and choose a corresponding value of $\phi_{0}$ such that Eq.~(\ref{eq:delta phi0 cond})
holds. Then the sequence $v^{(n)}(\phi)$ converges as $n\rightarrow\infty$
at any fixed $\phi>\phi_{0}$.

\subsection{Summary of assumptions\label{sub:Summary-of-assumptions}}

In the preceding sections we have established that a general model
of $k$-inflation admits an attractor solution $v_{*}(\phi)$ under
the following (sufficient) conditions:

1) The function $g(v,\phi)$ defined by Eq.~(\ref{eq:g def}) has
an asymptotic expansion at $\phi\rightarrow\infty$ of the form~(\ref{eq:asympt of g}),
where $B_{1}(\phi)/B_{0}(\phi)\rightarrow0$ as $\phi\rightarrow\infty$.
This asymptotic expansion defines the functions $A_{0}(v)$, $A_{1}(v)$.

2) The function $A_{0}(v)$ has a unique root $v_{0}<0$ such that
$A_{0}^{\prime}(v_{0})>0$.

3) The function $A_{1}(v)$ is regular at $v_{0}$ and does not grow
faster than $A_{0}(v)$ for large $v$. 

4) The function $B_{0}(\phi)$ is such that $\int^{\infty}B_{0}(\phi)d\phi=\infty$
(the integral diverges).

We have shown that the attractor solution $v_{*}(\phi)$ approaches
$v_{0}$ as $\phi\rightarrow\infty$ and has a well-defined asymptotic
expansion for $\phi\rightarrow\infty$, which can be determined by
iterating Eq.~(\ref{eq:inverse iteration}) starting with $v(\phi)\equiv v_{0}$.
The first iteration yields the slow-roll approximation to the attractor,
which is the function $v_{sr}(\phi)$ defined by\begin{equation}
g(v_{sr}(\phi),\phi)=0.\end{equation}
Assuming that (generically) $A_{1}(v_{0})\neq0$, we can obtain the
following leading-order \emph{approximate} expression for the slow-roll
solution,\begin{equation}
v_{sr}(\phi)\approx v_{0}-\frac{A_{1}(v_{0})}{A_{0}^{\prime}(v_{0})}\frac{B_{1}(\phi)}{B_{0}(\phi)}.\label{eq:sr approx}\end{equation}

In addition to the above assumptions, the following two conditions
must be satisfied in order for the attractor solution to be physically
relevant: (i) The energy density, $\varepsilon(v,\phi)\equiv vp_{,v}-p$,
is positive for all $v$ near $v_{0}$ and does not approach zero
as $\phi\rightarrow\infty$. (ii) The speed of sound is real-valued,
$c_{s}^{2}>0$. The condition~(i) will be satisfied if $\varepsilon(v_{0},\phi)>0$
for all sufficiently large $\phi$. The condition~(ii) says that\begin{equation}
c_{s}^{2}=\frac{p_{,v}}{\varepsilon_{,v}}=\left.\frac{p_{,v}}{vp_{,vv}}\right|_{v=v_{0}}>0.\label{eq:cond ii}\end{equation}
An explicit form of the condition~(\ref{eq:cond ii}) for the Lagrangian~(\ref{eq:KEssense p})
is given in Eq.~(\ref{eq:cs cond 1}) below.

\subsection{Existence of attractors in $k$-inflation\label{sub:Existence-of-attractors}}

We now apply the preceding constructions to a model with the Lagrangian~(\ref{eq:KEssense p}).
We first need to analyze the asymptotic form of the equation of motion,
$v'=g(v,\phi)$, at $\phi\rightarrow\infty$. The function $g(v,\phi)$
is given by Eq.~(\ref{eq:g for KEssence}), which is in the form~(\ref{eq:asympt of g})
with only two terms,\begin{align}
g(v,\phi) & =A_{0}(v)B_{0}(\phi)+A_{1}(v)B_{1}(\phi),\\
A_{0}(v) & \equiv-3\kappa\frac{Q^{\prime}(v)\sqrt{\tilde{\varepsilon}(v)}}{vQ^{\prime\prime}(v)},\quad A_{1}(v)\equiv-\frac{\tilde{\varepsilon}(v)}{vQ^{\prime\prime}(v)},\label{eq:A1 def}\\
\tilde{\varepsilon}(v) & \equiv vQ^{\prime}(v)-Q(v);\\
B_{0}(\phi) & \equiv\sqrt{K(\phi)},\quad B_{1}(\phi)\equiv\frac{d}{d\phi}\left[\ln K(\phi)\right].\label{eq:B0B1 def}\end{align}
 The asymptotic behavior of the functions $B_{0}(\phi)$ and $B_{1}(\phi)$
at $\phi\rightarrow\infty$ may belong to one of the three cases:
(i) $B_{0}(\phi)\ll B_{1}(\phi)$, (ii) $B_{0}(\phi)\gg B_{1}(\phi)$,
and (iii) $B_{0}(\phi)\sim B_{1}(\phi)$. In the first case, the equation
of motion is dominated by $A_{1}(v)$ which cannot have roots $A_{1}(v_{0})=0$,
otherwise the energy density would become small on the attractor,
$\varepsilon(v_{*},\phi)\approx\varepsilon(v_{0},\phi)=0$ at $\phi\rightarrow\infty$,
and could not dominate the energy density of other matter. Hence,
in case (i) inflationary attractors do not exist. Case (ii) covers
all functions $K(\phi)$ that either grow with $\phi$, or tend to
a constant, or decay slower than $\phi^{-2}$ as $\phi\rightarrow\infty$.
Case (iii) corresponds to a specific family of Lagrangians with $K(\phi)\propto\phi^{-2}$,
which gives $B_{0}(\phi)\sim B_{1}(\phi)\propto\phi^{-1}$. After
a change of variable, $\phi\rightarrow\tilde{\phi}\equiv\ln\phi$,
this case is reduced to case (ii) with $B_{0}=1$ and $B_{1}=0$.
Therefore, we shall confine our attention to case (ii). The precise
conditions for that case are \begin{align}
\frac{B_{1}(\phi)}{B_{0}(\phi)} & \equiv\frac{K^{\prime}(\phi)}{K^{3/2}(\phi)}\rightarrow0\textrm{ as }\phi\rightarrow\infty;\label{eq:K cond 1}\\
\int^{\infty}B_{0}(\phi)d\phi & =\int^{\infty}\sqrt{K(\phi)}d\phi=\infty.\label{eq:K cond 2}\end{align}

Assuming a polynomial growth of $Q(v)\sim v^{n}$ for large $v$,
where $n>2$, we find that $A_{0}(v)$ grows faster than $A_{1}(v)$;
this verifies condition~3 in Sec.~\ref{sub:Summary-of-assumptions}.
Therefore, the term $A_{0}(v)B_{0}(\phi)$ dominates the equation
of motion for non-attractor solutions at large $\phi$ and, as shown
in previous sections, an attractor solution $v_{*}(\phi)$ must tend
to a root $v_{0}\neq0$ of $A_{0}(v)$ as $\phi\rightarrow\infty$.
The function $A_{0}(v)$ is expressed by Eq.~(\ref{eq:A1 def}).
Since, as we already found, $\tilde{\varepsilon}(v)$ must be nonzero
to drive inflation, it follows that $v_{0}$ must be a root of $Q'(v)$.
Let us assume that $v_{0}<0$ is such a root, $Q'(v_{0})=0$. It remains
to verify the other conditions for the existence of the attractor.

The energy density is positive on the attractor solution, $\varepsilon=-K(\phi)Q(v_{0})>0$,
if $K(\phi)>0$ and $Q(v_{0})<0$. Thus, the only admissible roots
of $Q'(v_{0})$ are such that $K(\phi)Q(v_{0})<0$. The condition
$A_{0}^{\prime}(v_{0})>0$ is satisfied since $v_{0}<0$, $\tilde{\varepsilon}>0$,
and thus\begin{equation}
A_{0}^{\prime}(v_{0})=-3\kappa\sqrt{\tilde{\varepsilon}(v_{0})}\frac{Q^{\prime\prime}(v_{0})}{v_{0}Q^{\prime\prime}(v_{0})}=-3\kappa\frac{\sqrt{\tilde{\varepsilon}}}{v_{0}}>0,\end{equation}
 where we have assumed that (generically) $Q''(v_{0})\neq0$. Finally,
the speed of sound must satisfy the condition~(\ref{eq:cond ii}),
\begin{equation}
c_{s}^{2}=\frac{p_{,v}}{vp_{,vv}}=\frac{Q'(v_{*})}{v_{*}Q''(v_{*})}\approx\frac{v_{*}(\phi)-v_{0}}{v_{0}}>0.\label{eq:cs cond 0}\end{equation}
 This condition effectively constrains the deviation of the attractor
solution $v_{*}(\phi)$ from $v=v_{0}=\textrm{const}$, requiring
that $v_{*}(t)<v_{0}$. Because of that, we have $Q'(v_{*})<0$ and
$g(v_{*},\phi)>0$. Then the equation of motion, $dv_{*}/d\phi=g(v_{*},\phi)$,
entails that the negative value $v_{*}(\phi)$ monotonically increases
(in the algebraic sense) as $\phi$ decreases, at least until the
slow-roll approximation $v_{*}(\phi)\approx v_{0}$ breaks down. Numerical
simulations for some model Lagrangians show that $v_{*}(\phi)$ is
typically monotonic in $\phi$ all the way until the end of inflation.

Using the approximate solution~(\ref{eq:sr approx}) and the fact
that $v_{0}<0$, we rewrite the condition~(\ref{eq:cs cond 0}) as\begin{equation}
c_{s}^{2}=\frac{1}{\left|v_{0}\right|}\frac{A_{1}(v_{0})}{A_{0}^{\prime}(v_{0})}\frac{B_{1}(\phi)}{B_{0}(\phi)}=\frac{\sqrt{\tilde{\varepsilon}}}{3\kappa\left|v_{0}\right|Q^{\prime\prime}(v_{0})}\frac{K^{\prime}(\phi)}{K^{3/2}(\phi)}>0.\label{eq:cs cond 1}\end{equation}
It follows that $c_{s}^{2}>0$ if $Q^{\prime\prime}(v_{0})K^{\prime}(\phi)>0$
for large $\phi$. We conclude that a model of $k$-inflation satisfying
the conditions listed in Sec.~\ref{sec:Main-results} will admit
an attractor solution with properties suitable for inflationary cosmology.

\subsection{Slow-roll condition\label{sub:Slow-roll-condition}}

In this section we show that attractor solutions in $k$-inflation
models satisfy the slow-roll condition for large $\phi$. The slow-roll
condition means that the change in the Hubble rate, $\Delta H\equiv\dot{H}\Delta t$,
is negligible during one Hubble time $\Delta t=H^{-1}$, i.e.\begin{equation}
H^{-1}\left|\dot{H}\right|\ll H.\end{equation}
This condition guarantees that the spacetime is locally sufficiently
close to de Sitter during inflation. 

It follows from Eq.~(\ref{eq:epsilon EOM}) that\begin{equation}
\left|\frac{1}{H}\frac{dH}{dt}\right|=\frac{1}{2}\left|\frac{1}{\varepsilon}\frac{d\varepsilon}{dt}\right|=\frac{3}{2}H\left|\frac{\varepsilon+p}{\varepsilon}\right|,\end{equation}
therefore the slow-roll condition with the Lagrangian~(\ref{eq:KEssense p})
is equivalent to $\left|\varepsilon+p\right|\ll\varepsilon$ or\begin{equation}
\left|\frac{vp_{,v}}{vp_{,v}-p}\right|=\left|\frac{vQ'(v)}{vQ'(v)-Q(v)}\right|\ll1.\end{equation}
 Since the attractor approaches a root $v_{0}$ of $Q'(v)$ as $\phi\rightarrow\infty$,
and since $Q(v_{0})<0$ by the condition of the positivity of the
energy density, the slow-roll condition will be satisfied for large
enough $\phi$. More precisely, $\phi$ must be large enough so that
\begin{equation}
\frac{v_{0}Q''(v_{0})}{Q(v_{0})}\left|v(\phi)-v_{0}\right|\ll1.\label{eq:slowroll cond}\end{equation}

The slow-roll condition is violated (and inflation ends) when $\ddot{a}=0$,
which is equivalent to $\dot{H}+H^{2}=0$ or $\varepsilon+3p=0$.
This occurs at a time $t$ when either $vQ'(v)+2Q(v)=0$ with $v\equiv\dot{\phi}(t)$,
or $K(\phi(t))=0$ (then $\varepsilon=p=0$ at the same time). Generically,
we may expect that the former condition does not take place; so the
end-of-inflation point is the root $\phi_{E}$ of $K(\phi_{E})=0$.
For instance, this is the case when $K(\phi)$ is monotonic and $Q(v)=-c_{1}v^{n_{1}}+c_{2}v^{n_{2}}$
with $c_{1}>0$, $c_{2}>0$, $n_{2}>n_{1}>1$.

\subsection{Asymptotics of attractors in $k$-inflation\label{sub:Asymptotics-of-attractors-in-k}}

Assuming that the conditions for the existence of the attractor are
met, let us now compute the asymptotic form of the attractor solution
for large $\phi$. All the functions describing the homogeneous cosmological
solutions ($\dot{\phi}$, $H$, $c_{s}$, etc.) can be expressed as
functions of $\phi$. Substituting Eqs.~(\ref{eq:A1 def})-(\ref{eq:B0B1 def})
into Eq.~(\ref{eq:sr approx}), we find\begin{align}
\dot{\phi} & \equiv v_{*}(\phi)\approx v_{0}-v_{1}(\phi),\\
v_{1}(\phi) & \equiv\frac{\sqrt{-Q(v_{0})}}{3\kappa Q^{\prime\prime}(v_{0})}\frac{K'(\phi)}{K^{3/2}(\phi)},\\
\phi(t) & \approx\phi_{0}+v_{0}t+\frac{2}{3\kappa Q^{\prime\prime}(v_{0})}\sqrt{\frac{-Q(v_{0})}{K(\phi_{0}+v_{0}t)}},\\
H(\phi) & \approx\kappa\sqrt{-K(\phi)Q(v_{0})}\left(1+\frac{v_{0}Q^{\prime\prime}(v_{0})}{2Q(v_{0})}v_{1}(\phi)\right),\label{eq:H on attractor}\\
\ln a(\phi) & =\int_{\phi_{0}}^{\phi}\!\frac{H}{v}d\phi\approx\frac{\kappa\sqrt{-Q(v_{0})}}{\left|v_{0}\right|}\int_{\phi}^{\phi_{0}}\negmedspace\sqrt{K(\phi)}d\phi,\\
c_{s}^{2}(\phi) & \approx-\frac{v_{1}(\phi)}{v_{0}}=\frac{\sqrt{-Q(v_{0})}}{3\kappa\left|v_{0}\right|Q^{\prime\prime}(v_{0})}\frac{K'(\phi)}{K^{3/2}(\phi)}.\label{eq:cs on attractor}\end{align}
The assumed conditions are\begin{align}
v_{0} & <0,\quad Q'(v_{0})=0,\quad Q(v_{0})<0,\label{eq:conditions for attractor}\\
K(\phi) & >0,\quad K'(\phi)Q''(v_{0})>0.\label{eq:conditions for attractor 2}\end{align}

Note that $c_{s}\ll1$ for large $\phi$, due to the condition~(\ref{eq:K cond 1}).
The small value of $c_{s}^{2}$ can be considered a {}``slow-roll
parameter,'' i.e.~a parameter describing the smallness of the deviation
from the exact de Sitter evolution. The slow-roll condition~(\ref{eq:slowroll cond})
holds if\begin{equation}
\frac{v_{0}^{2}Q''(v_{0})}{Q(v_{0})}c_{s}^{2}(\phi)=\frac{\left|v_{0}\right|}{3\kappa\sqrt{\left|Q(v_{0})\right|}}\frac{K'(\phi)}{K^{3/2}(\phi)}\ll1.\end{equation}
This inequality determines a model-dependent slow-roll range $\phi>\phi_{sr}$.
The slow-roll approximation becomes increasingly precise in the large
$\phi$ limit where $c_{s}\rightarrow0$.

\section{Magnitude of quantum fluctuations\label{sec:Magnitude-of-quantum}}

In this section we compute the magnitude of quantum fluctuations of
the field $\phi$. This calculation will yield the relevant kinetic
coefficients for the Fokker-Planck equation, which will be considered
in Sec.~\ref{sec:Analysis-of-Fokker-Planck}.

\subsection{Quantization of fluctuations}

We consider cosmological perturbations of the field $\phi$ on the
inflationary attractor solution in a spatially flat FRW universe.
We follow Ref.~\cite{Garriga:1999vw}, where the quantum theory of
perturbations for fields with noncanonical kinetic terms was developed.

Perturbations of the field $\phi$ give rise to scalar perturbations
of the metric. Their simultaneous dynamics is described by a single
scalar field $u$ with the (classical) equation of motion\begin{equation}
u^{\prime\prime}-c_{s}^{2}\Delta u-\frac{z^{\prime\prime}}{z}u=0,\end{equation}
where $c_{s}$ is the speed of sound, the prime denotes derivatives
with respect to the conformal time,\begin{equation}
u^{\prime}\equiv\frac{\partial u}{\partial\eta}\equiv a(t)\frac{\partial u}{\partial t},\end{equation}
 and the auxiliary function $z(t)$ is defined by\begin{equation}
z\equiv\frac{a\sqrt{\varepsilon+p}}{c_{s}H}.\end{equation}
The fluctuation $\delta\phi$ of the field $\phi$ and the Newtonian
gravitational potential $\Phi$ are expressed through $u$ (in the
longitudinal gauge) as~\cite{Garriga:1999vw}\begin{align}
\frac{\delta\phi}{\dot{\phi}} & =\frac{u}{zH}-\frac{\Phi}{H},\\
\frac{\Phi}{H} & =\frac{3\kappa^{2}}{2}z^{2}\Delta^{-1}\frac{d}{dt}\left(\frac{u}{z}\right).\end{align}

It is convenient to consider the Fourier modes $\delta\phi_{k}$,
$\Phi_{k}$, $u_{k}$ of the perturbation variables; then the inverse
Laplace operator in the above equation becomes simply $-k^{-2}$.
For a model with the Lagrangian~(\ref{eq:KEssense p}) and the attractor
solution found in Sec.~\ref{sub:Asymptotics-of-attractors-in-k},
we have\begin{align}
z & =\frac{1}{\kappa}\left|v_{0}\right|a(t)\sqrt{\frac{Q_{0}^{\prime\prime}}{\left|Q_{0}\right|}}\left(1+O(c_{s}^{2})\right),\\
zH & =\left|v_{0}\right|a(t)\sqrt{K(\phi)Q_{0}^{\prime\prime}}\left(1+O(c_{s}^{2})\right),\end{align}
where we denoted for brevity $Q_{0}\equiv Q(v_{0})$, $Q_{0}^{\prime\prime}\equiv Q^{\prime\prime}(v_{0})$.
We find that $z(t)\approx a(t)\cdot\textrm{const}$; here and below
we denote by {}``$\approx$'' the approximation obtained by neglecting
terms of order $c_{s}^{2}\ll1$. Within this approximation, we can
express the field fluctuation $\delta\phi_{k}$ as\begin{align}
\delta\phi_{k} & =\frac{v_{0}}{zH}u_{k}-\frac{3\kappa^{2}}{2}v_{0}z^{2}\Delta^{-1}\frac{d}{dt}\left(\frac{u_{k}}{z}\right)\nonumber \\
 & \approx-\frac{u_{k}}{a\sqrt{KQ_{0}^{\prime\prime}}}-\frac{3\kappa a^{2}v_{0}^{2}}{2k^{2}}\sqrt{\frac{Q_{0}^{\prime\prime}}{\left|Q_{0}\right|}}\frac{d}{dt}\left(\frac{u_{k}}{a}\right).\label{eq:delta phi thru u}\end{align}

It is known~\cite{Garriga:1999vw} that the action functional for
the field $u(\mathbf{x},\eta)$ has the canonical form. Therefore,
one quantizes the field $u(\mathbf{x},\eta)$ by postulating the mode
expansion\begin{equation}
\hat{u}(\mathbf{x},\eta)=\int\frac{d^{3}\mathbf{k}}{\left(2\pi\right)^{3/2}}\frac{1}{\sqrt{2}}\left[\hat{a}_{\mathbf{k}}u_{k}^{*}(\eta)e^{i\mathbf{k}\cdot\mathbf{x}}+H.c.\right],\end{equation}
where $\hat{a}_{\mathbf{k}}$ is a canonical annihilation operator
satisfying $[\hat{a}_{\mathbf{k}},\hat{a}_{\mathbf{k}'}^{\dagger}]=\delta(\mathbf{k}-\mathbf{k}')$,
and {}``$H.c.$'' denotes Hermitian conjugate terms. The mode functions
$u_{k}(\eta)$ are normalized by the condition $\textrm{Im}(\dot{u}_{k}u_{k}^{*})=1$
and satisfy \begin{equation}
u_{k}^{\prime\prime}+c_{s}^{2}k^{2}u_{k}-\frac{a^{\prime\prime}}{a}u_{k}=0,\end{equation}
where we replaced $z''/z$ by $a''/a$, which is justified within
the slow-roll range. The conformal time variable $\eta$ can be expressed
as a function of $\phi$,\begin{align}
\eta & =-\int_{0}^{\phi}\frac{d\phi}{\left|v_{0}\right|a(\phi)}\nonumber \\
 & \approx-\int_{0}^{\phi}\frac{d\phi}{\left|v_{0}\right|}\exp\left[\frac{\kappa\sqrt{\left|Q_{0}\right|}}{\left|v_{0}\right|}\int^{\phi}\negmedspace\sqrt{K(\phi')}d\phi'\right]\nonumber \\
 & \approx-\frac{1}{\kappa\sqrt{\left|Q_{0}\right|K(\phi)}}\exp\left[\frac{\kappa\sqrt{\left|Q_{0}\right|}}{\left|v_{0}\right|}\int^{\phi}\negmedspace\sqrt{K(\phi')}d\phi'\right].\end{align}
 It follows that $a(\eta)\approx(H\eta)^{-1}$ and $a''/a\approx-2\eta^{-2}$,
so the mode functions $u_{k}(\eta)$ can be chosen as the standard
Bunch-Davies mode functions for the massless field (except for the
extra factors $c_{s}$),\begin{equation}
u_{k}(\eta)=\frac{1}{\sqrt{c_{s}k}}e^{ic_{s}k\eta}\left(1+\frac{i}{c_{s}k\eta}\right).\end{equation}

Since the field modes $\delta\phi_{k}$ are linearly related to $u_{k}$,
a similar mode expansion holds for the quantum field $\delta\hat{\phi}(\mathbf{x},t)$,
except for different mode functions as given by Eq.~(\ref{eq:delta phi thru u}):\begin{align}
\delta\hat{\phi}(\mathbf{x},\eta) & =\int\frac{d^{3}\mathbf{k}}{\left(2\pi\right)^{3/2}}\frac{1}{\sqrt{2}}\left[\hat{a}_{\mathbf{k}}\delta\phi_{k}(\eta)e^{i\mathbf{k}\cdot\mathbf{x}}+H.c.\right],\\
\delta\phi_{k}(\eta) & =-\frac{u_{k}}{a\sqrt{K(\phi)Q_{0}^{\prime\prime}}}-\frac{3\kappa a^{2}v_{0}^{2}}{2k^{2}}\sqrt{\frac{Q_{0}^{\prime\prime}}{\left|Q_{0}\right|}}\frac{d}{dt}\left(\frac{u_{k}}{a}\right)\nonumber \\
 & \approx\kappa\eta u_{k}(\eta)\left[\sqrt{\frac{\left|Q_{0}\right|}{Q_{0}^{\prime\prime}}}+\frac{3}{2}\frac{v_{0}^{2}c_{s}^{2}}{1-ic_{s}k\eta}\sqrt{\frac{Q_{0}^{\prime\prime}}{\left|Q_{0}\right|}}\right].\end{align}
We may now disregard the term of order $c_{s}^{2}$ and obtain\begin{equation}
\delta\phi_{k}(\eta)\approx\kappa\eta u_{k}(\eta)\sqrt{\frac{\left|Q_{0}\right|}{Q_{0}^{\prime\prime}}}=\kappa\frac{\eta e^{ic_{s}k\eta}}{\sqrt{c_{s}k}}\left[1+\frac{i}{c_{s}k\eta}\right]\negmedspace\negmedspace\sqrt{\frac{\left|Q_{0}\right|}{Q_{0}^{\prime\prime}}}.\label{eq:delta phi ans 1}\end{equation}
We conclude that in the slow-roll regime the mode function $\delta\phi_{k}(\eta)$
is proportional to the standard Bunch-Davies mode function evaluated
at the wavenumber $c_{s}k$.

\subsection{Scales of averaging\label{sub:Scales-of-averaging}}

We are now interested in the spatially averaged perturbations $\delta\bar{\phi}$
of the field $\phi$. It is well known \cite{Starobinsky:1986fx}
that quantum fluctuations of a scalar field (in the Bunch-Davies vacuum
state) in de Sitter spacetime become uncorrelated on Hubble distance
and time scales, $L\sim\delta t\sim H^{-1}$, and are transformed
into classical perturbations. In the present case, the relevant distance
scale is the sound horizon scale, $L\sim c_{s}H^{-1}$, while the
time scale remains unchanged, $\delta t\sim H^{-1}$. Although this
was already noted e.g.~in Ref.~\cite{Garriga:1999vw}, we shall
now justify this statement more formally using arguments similar to
those of Ref.~\cite{Winitzki:1999ve}. 

The averaged fluctuation field $\delta\bar{\phi}$ can be defined
as\begin{equation}
\delta\bar{\phi}(\mathbf{x},t)=\int d^{3}\mathbf{x}'\,\delta\hat{\phi}(\mathbf{x}',t)W(L^{-1}|\mathbf{x}-\mathbf{x}'|),\end{equation}
where $L$ is the averaging scale and $W(q)$ is a suitable window
function that quickly decays for $q\gtrsim1$. Then the effective
noise field is defined as \begin{equation}
\bar{\xi}\equiv\frac{d}{dt}\delta\bar{\phi}.\end{equation}
For a minimally coupled, massless scalar field $\phi$, it was shown
in Ref.~\cite{Winitzki:1999ve} that correlators of the noise field
$\bar{\xi}$ have the following large-distance asymptotics, which
hold for a wide class of window functions $W$:\begin{equation}
\left\langle \bar{\xi}(\mathbf{x},t)\bar{\xi}(\mathbf{x}',t')\right\rangle \sim\frac{e^{-2H\left|t-t'\right|}}{H^{4}\left|\mathbf{x}-\mathbf{x}'\right|^{4}}.\end{equation}
This indicates a quick decay on time and distance scales of order
the Hubble horizon. In the present calculation, the field $\delta\hat{\phi}$
differs from the minimally coupled massless field considered in Ref.~\cite{Winitzki:1999ve}
only by the replacement $k\rightarrow c_{s}k$ and by an overall factor;
this is easy to see by examining the mode function~(\ref{eq:delta phi ans 1}).
Therefore, the calculations of Ref.~\cite{Winitzki:1999ve} apply
also to the present case if we rescale the length as $L\rightarrow c_{s}^{-1}L$.
The time scale remains unchanged, whereas the distance scale is found
from the relation $c_{s}^{-1}L=H^{-1}$, which yields $L=c_{s}H^{-1}$.

\subsection{Diffusion coefficient\label{sub:Diffusion-coefficient}}

The effective dynamics of the averaged field $\bar{\phi}$ is described
by a Langevin equation,\begin{equation}
\frac{d\bar{\phi}}{dt}=v_{*}(\bar{\phi})+\bar{\xi}(\mathbf{x},t),\end{equation}
where $\bar{\xi}(\mathbf{x},t)$ is the effective {}``noise'' field,
which encapsulates the effects of subhorizon quantum fluctuations.
In the stochastic formalism, the {}``noise'' is treated as a \emph{classical}
field, i.e.~a Gaussian random function with known correlations ({}``colored
noise'')~\cite{Winitzki:1999ve,Matarrese:2003ye,Liguori:2004fa}.
It was shown in Sec.~\ref{sub:Scales-of-averaging} that the field
$\bar{\xi}$ is essentially uncorrelated beyond Hubble time scales
$\delta t\sim H$ and beyond distance scales $L\sim c_{s}H^{-1}$.
To simplify the analysis, one may treat the {}``noise'' field $\bar{\xi}$
approximately as white noise as long as one considers variations of
$\bar{\phi}$ only on sufficiently large distance and time scales.
Such variations of $\bar{\phi}$ can then be described by the following
Langevin equation, which is this time a \emph{difference} equation,\begin{equation}
\bar{\phi}(\mathbf{x},t+\delta t)-\bar{\phi}(\mathbf{x},t)=v_{*}(\bar{\phi})\delta t+\sqrt{2D(\bar{\phi})\delta t}\delta\gamma,\label{eq:Langevin fd}\end{equation}
where $D$ is a {}``diffusion coefficient'' and $\delta\gamma$
is a fiducial Gaussian random variable with unit variance. The coefficient
$D(\phi)$ is defined through the standard deviation of the quantum
{}``jump'' in the field $\bar{\phi}$ during a time interval $\delta t$,\begin{equation}
2D\delta t\equiv\left\langle \delta\bar{\phi}^{2}(\mathbf{x},t)\right\rangle _{L\sim c_{s}H^{-1}}\equiv\left\langle \left(\frac{d}{dt}\delta\bar{\hat{\phi}}\cdot\delta t\right)^{2}\right\rangle _{L\sim c_{s}H^{-1}},\label{eq:D coef def}\end{equation}
where we explicitly indicate averaging the field $\frac{d}{dt}\delta\hat{\phi}(\mathbf{x},t)$
on distance scales $L\sim c_{s}H^{-1}$. 

As long as one considers variations of $\bar{\phi}$ on scales at
least $\sim H^{-1}$, the difference equation~(\ref{eq:Langevin fd})
is equivalent to the stochastic differential equation\begin{equation}
\frac{d\bar{\phi}}{dt}=v_{*}(\phi)+\sqrt{2D(\phi)}\gamma,\end{equation}
where $\gamma$ is a fiducial {}``white noise'' field. The above
equation is, in turn, equivalent to a FP equation for the time-dependent
probability distribution $P_{c}(\phi,t)$, \begin{equation}
\frac{\partial P_{c}}{\partial t}=\frac{\partial^{2}}{\partial\phi^{2}}\left(DP_{c}\right)-\frac{\partial}{\partial\phi}\left(v_{*}P_{c}\right)\equiv\hat{L}_{c}P,\label{eq:FP equ c1}\end{equation}
where we use the Ito factor ordering in the FP equation~\cite{Vilenkin:1999kd}. 

We can now compute the diffusion coefficient $D(\phi)$ in the presently
considered model of $k$-inflation, using Eq.~(\ref{eq:D coef def}).
Field fluctuations on distance scales $L\sim c_{s}H^{-1}$ in the
Bunch-Davies vacuum state are due to field modes exiting the sound
horizon. Therefore, the magnitude of these fluctuations can be estimated
by evaluating the absolute value of the mode function~(\ref{eq:delta phi ans 1})
at the wavenumber $k$ that corresponds to the sound horizon crossing,
$c_{s}k=aH\approx\left|\eta\right|^{-1}$. Since the quantum field
$\frac{d}{dt}\delta\hat{\phi}$ has the standard mode expansion with
mode functions $\frac{d}{dt}\delta\phi_{k}$, we find using Eq.~(\ref{eq:delta phi ans 1})
that\begin{align}
\frac{d}{dt}\delta\phi_{k} & =a^{-1}\frac{d}{d\eta}\kappa\frac{\eta e^{ic_{s}k\eta}}{\sqrt{c_{s}k}}\left[1+\frac{i}{c_{s}k\eta}\right]\sqrt{\frac{\left|Q_{0}\right|}{Q_{0}^{\prime\prime}}}\\
 & =\kappa\sqrt{\frac{\left|Q_{0}\right|}{Q_{0}^{\prime\prime}}}\frac{e^{ic_{s}k\eta}}{\sqrt{c_{s}k}}ic_{s}k\eta^{2}H.\end{align}
 Therefore, the typical fluctuation in $\frac{d}{dt}\delta\hat{\phi}$
averaged on scales $L\sim c_{s}H^{-1}$ and accumulated during one
Hubble time, $\delta t=H^{-1}$, is \begin{align}
\left\langle \delta\bar{\phi}^{2}\delta t^{2}\right\rangle _{L\sim c_{s}H^{-1}} & \approx\left.\left(\frac{k^{3}}{4\pi^{2}}\left|\frac{d}{dt}\delta\phi_{k}\right|^{2}H^{-2}\right)\right|_{c_{s}k\left|\eta\right|=1}\nonumber \\
 & =\frac{\kappa^{2}}{4\pi^{2}}\frac{\left|Q_{0}\right|}{Q_{0}^{\prime\prime}}c_{s}^{-3}.\end{align}
Finally, the coefficient $D(\phi)$ is computed from Eq.~(\ref{eq:D coef def}),
\begin{equation}
D=\frac{1}{2\delta t}\left\langle \left(\frac{d}{dt}\delta\bar{\phi}\cdot\delta t\right)^{2}\right\rangle _{L\sim c_{s}H^{-1}}\approx\frac{\kappa^{2}}{8\pi^{2}}\frac{\left|Q_{0}\right|}{Q_{0}^{\prime\prime}}\frac{H}{c_{s}^{3}}.\label{eq:D coef ans}\end{equation}
Since $c_{s}\ll1$ for most of the allowed range of $\phi$, the diffusion
coefficient is typically large. As we shall see, this leads to the
presence of self-reproduction.

A more explicit expression for $D(\phi)$ can be found using Eqs.~(\ref{eq:H on attractor})
and (\ref{eq:cs on attractor}),\begin{equation}
D(\phi)=\frac{\kappa^{9/2}\left|Q_{0}\right|^{3/4}\left|v_{0}\right|^{3/2}\sqrt{27Q_{0}^{\prime\prime}}}{8\pi^{2}}\frac{\left[K(\phi)\right]^{11/4}}{\left[K'(\phi)\right]^{3/2}}.\label{eq:D explicit}\end{equation}

\subsection{Conditions for self-reproduction\label{sub:Conditions-for-self-reproduction}}

Qualitatively, self-reproduction is present if the typical change
in the averaged field due to random quantum jumps ({}``diffusion'')
during one Hubble time is much larger than the deterministic change
due to the classical equation of motion,\begin{equation}
\sqrt{2D(\phi)\delta t}\gg v(\phi)\delta t,\quad\delta t\equiv H^{-1}.\end{equation}
It is straightforward to see that this condition generically holds
in the slow-roll regime. Using Eq.~(\ref{eq:D coef ans}), we find\begin{equation}
\frac{\left(v\delta t\right)^{2}}{D\delta t}\sim\frac{v_{0}^{2}}{DH}\sim\frac{c_{s}^{3}}{H^{2}}\rightarrow0\quad\textrm{as }\phi\rightarrow\infty,\end{equation}
since $c_{s}(\phi)\rightarrow0$, whereas $H(\phi)\sim\sqrt{K(\phi)}$
remains constant or grows as $\phi\rightarrow\infty$. Therefore,
diffusion will dominate over the deterministic drift for all sufficiently
large $\phi$. Moreover, since the slow-roll condition is essentially
the same as $c_{s}^{2}\ll1$, diffusion becomes negligible only near
the end of the slow-roll range or even beyond that, near the end of
inflation. A wide range of $\phi$ where diffusion dominates over
the deterministic drift is a qualitative indication of eternal self-reproduction.
A precise criterion for the presence of self-reproduction is provided
by the Fokker-Planck equation for the volume-weighted distribution
of the field, which will be analyzed in Sec.~\ref{sec:Analysis-of-Fokker-Planck}.

\subsection{Validity of diffusion approximation\label{sub:Validity-of-diffusion}}

We have computed the magnitude $\delta\phi$ of a typical random {}``jump'',
defined as $\delta\phi\sim\sqrt{2D(\phi)\delta t}$, during one Hubble
time $\delta t=H^{-1}$. The diffusion approximation is valid if $\delta\phi$
is {}``small'' in the sense that it causes small changes in the
energy density of the field, \begin{equation}
\delta\varepsilon\equiv\varepsilon_{,\phi}\delta\phi\ll\varepsilon(\phi).\end{equation}
 If this condition does not hold, we would not be justified in assuming
that the {}``jumps'' cause a perturbatively small change in the
local expansion rate $H$. Since $\varepsilon_{,\phi}/\varepsilon\sim H_{,\phi}/H$,
an equivalent condition is \begin{equation}
\frac{H_{\phi}}{H}\delta\phi\equiv\frac{H_{\phi}}{H}\sqrt{2D(\phi)H^{-1}}\ll1.\label{eq:diff conditions}\end{equation}

There is another aspect that limits the validity of the diffusion
approach. As we have seen, the {}``jumps'' are uncorrelated on distances
of order $c_{s}H^{-1}$, which is significantly smaller than the Hubble
scale $H^{-1}$ since $c_{s}\ll1$. A comoving scale $c_{s}H^{-1}$
grows to Hubble size within the time $\delta t_{g}=H^{-1}\ln c_{s}^{-1}$.
Thus, a single Hubble-sized region will contain small inhomogeneities
in the values of $\phi$ on scales above $c_{s}H^{-1}$, generated
during the time interval $\delta t_{g}\gtrsim H^{-1}$. The typical
magnitude of these inhomogeneities,\begin{equation}
\delta\phi_{g}\sim\sqrt{2D(\phi)\delta t_{g}}=\sqrt{2DH^{-1}\ln c_{s}^{-1}},\end{equation}
 should not lead to a significant variation of $H$ within a Hubble-sized
region. (Otherwise, we cannot approximate the spacetime as a locally
homogeneous de Sitter with a well-defined value of $H$.) Thus we
obtain the condition\begin{equation}
\frac{H_{,\phi}}{H}\delta\phi_{g}\equiv\frac{H_{,\phi}}{H}\sqrt{\ln c_{s}^{-1}}\delta\phi=\sqrt{2DH^{-1}\ln c_{s}^{-1}}\frac{H_{,\phi}}{H}\ll1,\end{equation}
which is stronger than the condition~(\ref{eq:diff conditions})
by the factor $\sqrt{\ln c_{s}^{-1}}$. However, this factor is an
extremely slowly growing function of $\phi$, which we may expect
to contribute at most a factor of 2 or 3, and definitely no more than
one order of magnitude. Therefore, we may ignore this factor and concentrate
on the condition~(\ref{eq:diff conditions}).

To investigate whether the condition~(\ref{eq:diff conditions})
may be violated at large $\phi$ where $v_{*}(\phi)\approx v_{0}$,
we can use the analytic approximations derived above. Using the formulas~(\ref{eq:H on attractor}),
(\ref{eq:cs on attractor}), and (\ref{eq:D coef ans}), we find\begin{align}
\delta\phi & \sim\sqrt{2DH^{-1}}=\frac{\kappa}{4\pi}\sqrt{\frac{-Q_{0}}{Q_{0}^{\prime\prime}}}c_{s}^{-3/2}=b_{1}\frac{K^{9/8}(\phi)}{\left(K_{,\phi}\right)^{3/4}},\\
b_{1} & \equiv\frac{\kappa^{7/4}}{4\pi}\left(-Q_{0}\right)^{1/8}\left(Q_{0}^{\prime\prime}\right)^{1/4}\left[3\left|v_{0}\right|\right]^{3/4}.\end{align}
The constant coefficient $b_{1}$ is small since it contains a large
power of the inverse Planck mass, while other relevant energy scales
are presumably much lower. Since $\left(\ln H\right)_{,\phi}=\left(\ln K\right)_{,\phi}$,
the condition~(\ref{eq:diff conditions}) yields\begin{equation}
b_{1}\left(K_{,\phi}\sqrt{K}\right)^{1/4}\ll1.\end{equation}
Assuming the asymptotic behavior $K(\phi)\sim\phi^{s}$ for large
$\phi$, where $s\geq0$, we simplify the above condition to\begin{equation}
b_{1}\phi^{\frac{3}{8}s-\frac{1}{4}}\ll1.\end{equation}
This condition is satisfied for large $\phi$ if $0\leq s\leq\frac{2}{3}$
but leads to an upper bound, $\phi<\phi_{\max}$, on values of $\phi$
if $s>\frac{2}{3}$. In the latter case, for $\phi>\phi_{\max}$ the
fluctuations are so large that a single {}``quantum jump'' (within
a Hubble time $\delta t\sim H^{-1}$) may bring the value of $\phi$
outside of the slow-roll range, e.g.~reach the end of inflation.
It is clear that the diffusion approximation breaks down for $\phi>\phi_{\max}$.
In other words, regions of the universe with $\phi>\phi_{\max}$ cannot
be described within the present semiclassical framework because quantum
fluctuations are too large. A boundary condition needs to be imposed
at $\phi=\phi_{\max}$ if we wish to proceed using the diffusion approach.
One may impose an absorbing boundary condition, arguing that regions
with $\phi>\phi_{\max}$ disappear into a {}``sea of eternal randomness.''
Alternatively, one may impose a reflecting boundary condition, arguing
that the {}``sea of randomness'' emits and absorbs equally many
regions. Since this boundary condition serves only to validate the
diffusion approach, results can be trusted only if they are insensitive
to the chosen type of the boundary condition.

Models with $K(\phi)\sim\phi^{s}$ and $0\leq s\leq\frac{2}{3}$ do
not exhibit the problem described above, however it is still necessary
to impose a boundary condition at some value $\phi=\phi_{\max}$.
For instance, the energy density for large $\phi$ may reach the Planck
value, which determines a (model-dependent) Planck boundary $\phi_{\max}$
such that $K(\phi_{\max})\left|Q_{0}\right|\sim M_{Pl}^{4}$. Results
of applying the diffusion approach should be insensitive to the type
of the boundary condition as well as to the precise value of $\phi_{\max}$.

\section{Analysis of Fokker-Planck equations\label{sec:Analysis-of-Fokker-Planck}}

\subsection{Self-adjoint form of the FP equation}

As is long known, the FP equation~(\ref{eq:FP equ}) can be reduced
to a manifestly self-adjoint form, which is formally similar to a
Schrödinger equation for a particle in a one-dimensional potential.
One can then use standard results about the existence of bound states,
which correspond to stationary solutions of the FP equation. 

A stationary solution of the FP equation is of the form\begin{equation}
P_{p}(\phi,t)=P_{(\lambda)}(\phi)e^{\lambda t},\end{equation}
 where $P_{(\lambda)}(\phi)$ is a stationary probability distribution.
The function $P_{(\lambda)}(\phi)$ must be everywhere positive, integrable,
and satisfy the stationary FP equation,\begin{equation}
\hat{L}_{p}P_{(\lambda)}\equiv\left(DP_{(\lambda)}\right)_{,\phi\phi}-\left(v_{*}P_{(\lambda)}\right)_{,\phi}+3HP_{(\lambda)}=\lambda P_{(\lambda)}.\label{eq:stationary FP 1}\end{equation}
The proper time $t$ may be replaced by another time variable, $d\tau=T(\phi)dt$,
where $T(\phi)$ is an arbitrary positive function of $\phi$, and
it is implied that the value of $\tau$ is obtained by integrating
$\int T(\phi)dt$ along comoving worldlines~\cite{Winitzki:1995pg}.
The FP equation in the time gauge $\tau$ has the same form as Eq.~(\ref{eq:stationary FP 1}),
except that the coefficients $D,v,H$ are divided by $T(\phi)$. For
instance, with $T(\phi)=H(\phi)$ we obtain the {}``scale factor''
or {}``$e$-folding'' time variable $\tau=\int Hdt=\ln a$, and
the stationary FP equation for $P_{(\lambda)}^{(sf)}(\phi)$is\begin{equation}
\left(\frac{D}{H}P_{(\lambda)}^{(sf)}\right)_{,\phi\phi}-\left(\frac{v_{*}}{H}P_{(\lambda)}^{(sf)}\right)_{,\phi}+3P_{(\lambda)}^{(sf)}=\lambda P_{(\lambda)}^{(sf)}.\end{equation}
 Self-reproduction is eternal if there exists a stationary solution
$P_{(\lambda)}$ of Eq.~(\ref{eq:stationary FP 1}) with a positive
$\lambda$. This condition is independent of the time gauge within
the class of gauges $d\tau=T(\phi)dt$~\cite{Winitzki:2001np}. 

To investigate the existence of positive eigenvalues of the operator
$\hat{L}_{p}$, it is convenient to transform the stationary FP equation
into an explicitly self-adjoint form. We perform the calculation following
Ref.~\cite{Winitzki:1995pg} and introduce new variables as follows,\begin{align}
x\equiv x(\phi) & \equiv\int^{\phi}\frac{d\phi}{\sqrt{D(\phi)}},\quad\partial_{\phi}=\frac{1}{\sqrt{D}}\partial_{x},\label{eq:x new def}\\
P_{(\lambda)}(\phi) & \equiv\psi(x)D^{-3/4}(\phi)\,\exp\left[\frac{1}{2}\int^{\phi}\frac{v_{*}(\phi)}{D(\phi)}d\phi\right],\label{eq:new P psi def}\end{align}
where the replacement $\phi\rightarrow\phi(x)$ is implied in the
last line. Note that $P_{\lambda}(\phi)$ is being multiplied by a
strictly positive function of $\phi$, so the positivity of $\psi(x)$
will entail the positivity of $P_{(\lambda)}(\phi)$. Under the replacements~(\ref{eq:x new def})-(\ref{eq:new P psi def}),
Eq.~(\ref{eq:stationary FP 1}) is transformed into\begin{equation}
\psi_{,xx}-U(x)\psi=\lambda\psi,\label{eq:psi FP}\end{equation}
where $U(x)$ is the {}``potential'' defined as a function of $\phi$
through the relation\begin{equation}
U(x)\equiv-3H(\phi)+\frac{3}{16}\frac{\left(D_{,\phi}\right)^{2}}{D}-\frac{D_{,\phi\phi}}{4}-\frac{v_{*}D_{,\phi}}{2D}+\frac{v_{*,\phi}}{2}+\frac{v_{*}^{2}}{4D}.\end{equation}
Equation~(\ref{eq:psi FP}) is the self-adjoint form of the stationary
FP equation and is formally equivalent to a stationary Schrödinger
equation for a particle in a one-dimensional {}``potential'' $U(x)$
with the {}``energy'' $E\equiv-\lambda$. It is well-known that
the ground state of a Schrödinger equation can be chosen as a nonnegative
wavefunction. If we show that Eq.~(\ref{eq:psi FP}) has a ground
state $\psi(x)$ with a negative value of {}``energy'' $E_{0}<0$,
it will follow that there exists a stationary probability distribution
$P_{(\lambda_{\max})}(\phi)$ corresponding to the largest eigenvalue
$\lambda_{\max}>0$ of the FP equation. Then the existence of eternal
self-reproduction will be established.

In order to find out whether the ground state energy in a given potential
$U(x)$ is negative, it is convenient to use the variational principle,\begin{equation}
E_{0}=\min_{\psi(x)}\frac{\int\bar{\psi}\left(-\psi_{,xx}+U\psi\right)dx}{\int\bar{\psi}\psi dx}.\label{eq:E0 var}\end{equation}
We may now substitute any test function $\psi(x)$ and obtain an upper
bound on $E_{0}$. If we find such $\psi(x)$ that the integral in
the numerator of Eq.~(\ref{eq:E0 var}) is negative, it will follow
that $E_{0}<0$. It is sufficient to consider real-valued $\psi(x)$
that vanish at the boundaries of the range of integration. Integrating
by parts and omitting boundary terms, we find\begin{equation}
\int\psi\left(-\psi_{,xx}+U\psi\right)dx=\int\left[\left(\psi_{,x}\right)^{2}+U\psi^{2}\right]dx.\end{equation}
If $U(x)$ is everywhere positive, then clearly $E_{0}>0$, so the
only possibility to have a negative $E_{0}$ is to have a range of
$x$ where $U(x)<0$. Let us suppose that $U(x)<0$ within a range
$x_{1}<x<x_{2}$, with a typical value $U(x)\sim-U_{0}<0$ in the
middle of the range. Then we may choose a test function $\psi(x)$
so that $\psi(x)=0$ outside of the range $\left[x_{1},x_{2}\right]$
and $\psi(x)\sim1$ within that range. The typical value of $\psi'(x)$
will be of order $\left(x_{2}-x_{1}\right)^{-1}\equiv l^{-1}$, and
so we obtain a bound \begin{equation}
E_{0}\leq\frac{\int\left(\psi^{\prime2}+U\psi^{2}\right)dx}{\int\psi^{2}dx}\sim\frac{l^{-1}-U_{0}l}{l}=-U_{0}+\frac{1}{l^{2}}.\end{equation}
The desired negative bound will be achieved if the width $l$ of the
interval is sufficiently large so that $l^{2}>U_{0}^{-1}$. However,
the width $l$ is constrained by\begin{equation}
l\leq\frac{U_{0}}{\max_{x\in\left[x_{1},x_{2}\right]}\left|U'(x)\right|},\end{equation}
 since the function $U(x)$ must vary between $0$ and $-U_{0}$ within
the interval of width $l$. Hence,\begin{equation}
E_{0}\leq-U_{0}+\left[\frac{\max_{x\in\left[x_{1},x_{2}\right]}\left|U'(x)\right|}{U_{0}}\right]^{2}.\label{eq:E0 cond 0}\end{equation}
In the next section, we shall use this condition near a point $x$
where $U(x)=-U_{0}<0$.

\subsection{Stationary solutions in $k$-inflation\label{sub:Stationary-solutions-in-k}}

Let us now analyze a model of $k$-inflation with the Lagrangian~(\ref{eq:KEssense p}),
where \begin{equation}
K(\phi)\sim\phi^{s},\quad s\geq0,\end{equation}
for large $\phi$. We assume that the conditions for the existence
of an inflationary attractor are met. Let us now estimate the {}``potential''
$U(x)$ at values of $x$ that correspond to large $\phi$. (Note
that an infinite range of $\phi$ may be mapped into a finite range
of $x$.) It is more convenient to analyze the behavior of the {}``potential''
$U$ as a function of $\phi$. We find the following asymptotic behavior
of parameters at $\phi\rightarrow\infty$:\begin{align}
H(\phi) & \sim\phi^{s/2},\quad v(\phi)\approx v_{0}=\textrm{const},\\
c_{s}(\phi) & \sim\phi^{-\frac{1}{2}-\frac{s}{4}},\\
D(\phi) & \sim\phi^{n(s)},\quad n(s)\equiv\frac{3}{2}+\frac{5}{4}s.\end{align}
The dominant negative term in $U(x)$ is $-3H(\phi)\sim-\phi^{s/2}$,
while the dominant positive term could be only\begin{equation}
\frac{3}{16}\frac{\left(D_{,\phi}\right)^{2}}{D}-\frac{D_{,\phi\phi}}{4}\sim n\left(4-n\right)\phi^{n-2}.\label{eq:term 2}\end{equation}
However, the condition~(\ref{eq:diff conditions}), rewritten as\begin{equation}
\sqrt{DH^{-1}}\frac{H_{,\phi}}{H}\ll1,\label{eq:diff ok}\end{equation}
 together with the power-law dependences of $D(\phi)$, $H(\phi)$,
yields\begin{align}
D_{,\phi\phi} & \sim\frac{D}{\phi^{2}},\quad H_{,\phi}\sim\frac{H}{\phi},\\
\frac{D_{,\phi\phi}}{H} & \sim\frac{D}{H}\frac{1}{\phi^{2}}\sim DH^{-1}\left(\frac{H_{,\phi}}{H}\right)^{2}\ll1.\label{eq:D phi less H}\end{align}
Since the condition~(\ref{eq:diff ok}) must be satisfied for the
entire range of $\phi$ under consideration, the term~(\ref{eq:term 2})
is negligible compared with $-3H$. Similarly, the other terms decay
as $\phi^{-1}$ or faster for large $\phi$; for instance, $v_{*}^{2}/D\ll H$
within the entire slow-roll regime (see Sec.~\ref{sub:Conditions-for-self-reproduction}).
Therefore, the potential $U(x)$ can be estimated as $U(x)\approx-3H(\phi)$
for most of the allowed range of $\phi$, excluding only a narrow
range near the end of inflation. 

We now use the condition~(\ref{eq:E0 cond 0}) to show that the ground
state energy corresponding to the potential $U(x)$ is negative. Since
$dx/d\phi>0$, and since the function $H(\phi)\sim\phi^{s/2}$ grows
monotonically, the largest value of $H_{,x}$ is at the largest allowed
$\phi$. The bound~(\ref{eq:E0 cond 0}) applied to an interval $\left[\phi_{1},\phi_{2}\right]$,
upon using $dx=\sqrt{D}d\phi$, yields\begin{align}
E_{0} & <-3H(\phi_{2})+D(\phi_{2})\left(\frac{H_{,\phi}}{H}\right)^{2}\nonumber \\
 & \approx-3H(\phi_{2})+D(\phi_{2})\phi_{2}^{-2}\approx-3H(\phi_{2})<0,\end{align}
since $D\phi^{-2}\ll H$ according to Eq.~(\ref{eq:D phi less H}).
Thus we have shown that the ground state has negative {}``energy.''
This proves that the largest eigenvalue $\lambda_{\max}$ of the operator
$\hat{L}_{p}$ is positive, and hence that eternal self-reproduction
is present.

The bound $E_{0}<-3H(\phi_{2})$, where $\phi_{2}$ is essentially
any value within the allowed range $\phi_{sr}<\phi<\phi_{\max}$,
means that the largest eigenvalue $\lambda_{\max}$ is approximately
equal to $3H_{\max}$, where $H_{\max}$ is the largest accessible
value of $H(\phi)$. It follows that the fractal dimension of the
inflating domain (defined in a gauge-invariant way in Ref.~\cite{Winitzki:2001np})
is generically close to 3.

Finally, we note that the potential $U(x)\approx-3H(\phi)$ has a
global minimum near the upper boundary $\phi_{\max}$. It follows
that the ground state $\psi(x)$ will have a sharp maximum near the
upper boundary. Therefore, the stationary distribution $P_{(\lambda_{\max})}(\phi)$
indicates that most of the 3-volume contains $\phi\sim\phi_{\max}$.
However, it has been long known that that the distributions $P_{c}(\phi,t)$
and $P_{p}(\phi,t)$ are sensitive to the choice of equal-time hypersurfaces,
or the {}``time gauge'' (see e.g.~\cite{Garcia-Bellido:1993wn,Linde:1993xx,Linde:1994gy,Winitzki:1995pg,Linde:1995uf,Vilenkin:1998kr}).
For instance, the 3-volume may be dominated by $\phi\sim\phi_{\max}$
in one gauge but not in another; even the exponential growth of the
3-volume of the inflating domain is gauge-dependent~\cite{Winitzki:2005ya}.
Nevertheless, the distributions $P_{c}(\phi,t)$ and $P_{p}(\phi,t)$
provide a useful qualitative picture of the global features of the
spacetime during eternal inflation.

\subsection{Eigenvalues of the comoving FP equation\label{sub:Eigenvalues-of-the-comoving-FP}}

We now consider the comoving distribution $P_{c}(\phi,t)$, which
describes the values of $\phi$ along a single, randomly chosen comoving
worldline. This distribution satisfies Eq.~(\ref{eq:FP equ c}),\begin{equation}
\frac{dP_{c}}{dt}=\hat{L}_{c}P_{c},\label{eq:Pc equ}\end{equation}
 which can be brought into a self-adjoint form similarly to the proper-volume
FP equation. However, let us show directly that all the eigenvalues
of the operator $\hat{L}_{c}$ (with appropriate boundary conditions)
are negative. It will follow that the late-time asymptotic of Eq.~(\ref{eq:Pc equ})
is\begin{equation}
P_{c}(\phi,t)=P_{(\lambda_{0})}(\phi)e^{\lambda_{0}t},\quad\lambda_{0}<0.\end{equation}
In other words, the probability $P_{c}(\phi,t)$ tends uniformly to
zero at late times, which is interpreted to mean that the evolution
of $\phi$ along any particular comoving worldline will almost surely
(with probability 1) eventually arrive at one of the boundaries, either
at $\phi_{E}$ or at $\phi_{\max}$.

Equation~(\ref{eq:Pc equ}) can be written as a {}``conservation
law,''\begin{equation}
\partial_{t}P_{c}=\partial_{\phi}J,\quad J(\phi)\equiv\partial_{\phi}(DP_{c})-v_{*}P_{c},\label{eq:cons law}\end{equation}
where $J$ plays the role of the {}``current.'' The relevant types
of boundary conditions, which were discussed above, are conveniently
expressed in terms of the quantity $J$. Namely, the reflecting condition
is $J=0$, the absorbing condition is $P_{c}=0$, and the {}``exit-only''
condition (to be imposed at the end-of-inflation point $\phi_{E}$)
is $J=-v_{*}P_{c}$, meaning that diffusion cannot bring the field
from $\phi=\phi_{E}$ back into the inflationary range. To avoid considering
the absorbing and the reflecting conditions separately, we shall impose
at $\phi=\phi_{\max}$ a formal combination of an absorbing and reflecting
condition, $J+\alpha P_{c}=0$, where $\alpha\geq0$. Suppose $P_{(\lambda)}(\phi)$
is an eigenfunction satisfying $\hat{L}_{c}P_{(\lambda)}=\lambda P_{(\lambda)}$
with the boundary conditions\begin{align}
\left.\left(J_{(\lambda)}+v_{*}P_{(\lambda)}\right)\right|_{\phi_{E}} & =0,\quad\left.\left(J_{(\lambda)}+\alpha P_{(\lambda)}\right)\right|_{\phi_{\max}}=0,\label{eq:bc J P}\\
J_{(\lambda)} & \equiv\partial_{\phi}(DP_{(\lambda)})-v_{*}P_{(\lambda)}.\end{align}
 Note that the operator $\hat{L}_{c}$ is self-adjoint with respect
to the scalar product in the {}``$\psi$'' space,\begin{align}
\left(P_{1},P_{2}\right) & \equiv\int\psi_{1}(x)\psi_{2}(x)dx=\int P_{1}(\phi)P_{2}(\phi)M(\phi),\nonumber \\
M(\phi) & \equiv D(\phi)\exp\left[-\int^{\phi}\frac{v_{*}(\phi)}{D(\phi)}d\phi\right].\end{align}
So let us consider the scalar product of $P_{(\lambda)}$ and $\hat{L}_{c}P_{(\lambda)}$,
\begin{equation}
I[P_{(\lambda)}]\equiv\left(P_{(\lambda)},\hat{L}_{c}P_{(\lambda)}\right)=\int_{\phi_{E}}^{\phi_{\max}}P_{(\lambda)}\left(\hat{L}_{c}P_{(\lambda)}\right)Md\phi.\label{eq:I int}\end{equation}
By assumption, $\hat{L}_{c}P_{(\lambda)}=\lambda P_{(\lambda)}$,
thus\begin{equation}
I[P_{(\lambda)}]=\lambda\int_{\phi_{E}}^{\phi_{\max}}P_{(\lambda)}^{2}Md\phi.\end{equation}
Since $M(\phi)>0$, we will prove that $\lambda<0$ if we show that
$I[P_{(\lambda)}]<0$. 

Writing $\hat{L}_{c}P_{(\lambda)}=\partial_{\phi}J_{(\lambda)}$ and
integrating Eq.~(\ref{eq:I int}) by parts, we find\begin{equation}
I[P_{(\lambda)}]=\left.MP_{(\lambda)}J_{(\lambda)}\right|_{\phi_{E}}^{\phi_{\max}}-\int_{\phi_{E}}^{\phi_{\max}}d\phi\, J_{(\lambda)}\partial_{\phi}(MP_{(\lambda)}).\label{eq:I interm 1}\end{equation}
Using the boundary conditions~(\ref{eq:bc J P}), we can estimate
the boundary term $MP_{(\lambda)}J_{(\lambda)}$ as follows,\begin{align}
 & \left.\left(MP_{(\lambda)}J_{(\lambda)}\right)\right|_{\phi_{E}}^{\phi_{\max}}\nonumber \\
 & \;=-\alpha MP_{(\lambda)}^{2}(\phi_{\max})+v_{*}MP_{(\lambda)}^{2}(\phi_{E})<0\end{align}
because $v_{*}(\phi)<0$ and $\alpha\geq0$. It remains to put a bound
on the integral in Eq.~(\ref{eq:I interm 1}). It is easy to see
that $\partial_{\phi}(MP_{(\lambda)})$ is proportional to $J_{(\lambda)}$,
namely\begin{equation}
\partial_{\phi}(MP_{(\lambda)})=\frac{M(\phi)}{D(\phi)}J_{(\lambda)}(\phi).\end{equation}
 Since $M(\phi)/D(\phi)>0$, the integral in Eq.~(\ref{eq:I interm 1})
must be negative (note that $J_{(\lambda)}(\phi)$ is not everywhere
zero):\begin{equation}
-\int_{\phi_{E}}^{\phi_{\max}}d\phi\, J_{(\lambda)}\partial_{\phi}(MP_{(\lambda)})=-\int_{\phi_{E}}^{\phi_{\max}}d\phi\, J_{(\lambda)}^{2}\frac{M}{D}<0.\end{equation}
Thus we have shown that $I[P_{(\lambda)}]<0$, which proves that every
eigenvalue $\lambda$ of $\hat{L}_{c}$ is negative.

\subsection{Exit probability in $k$-inflation\label{sub:Exit-probability-in}}

We have seen that the evolution of the averaged field $\bar{\phi}$
is heavily influenced by random {}``jumps,'' which can even bring
the value of $\bar{\phi}$ to the upper boundary $\phi_{\max}$ where
the semiclassical description breaks down. Thus, we may trace the
evolution of $\bar{\phi}$ along a single comoving trajectory, starting
from an initial value $\phi_{0}$, and ask for the probability $p_{\textrm{exit}}$
of eventually exiting through the end-of-inflation boundary $\phi_{E}$
while staying away from the upper boundary $\phi_{\max}$. In this
section, we show that this exit probability is approximately equal
to 1 if $\phi_{0}$ is sufficiently far away from $\phi_{\max}$.

The exit probability can be found as follows (a similar method was
used in Ref.~\cite{Vilenkin:1999kd}). Let us assume that the initial
distribution is concentrated at $\phi=\phi_{0}$, i.e.\begin{equation}
P_{c}(\phi,t=0)=\delta(\phi-\phi_{0}),\end{equation}
where $\phi_{E}<\phi_{0}<\phi_{\max}$, and that the distribution
$P_{c}(\phi,t)$ is known at all times. Due to the conservation law~(\ref{eq:cons law}),
the probability of exiting inflation through $\phi=\phi_{E}$ during
a time interval $[t,t+dt]$ is\begin{align}
dp_{\textrm{exit}} & =Jdt\equiv\left.\left(\partial_{\phi}\left(DP_{c}\right)-v_{*}P_{c}\right)\right|_{\phi=\phi_{E}}dt\nonumber \\
 & =-v_{*}(\phi_{E})P_{c}(\phi_{E},t)dt\end{align}
(note that $v_{*}(\phi_{E})<0$), hence the total probability of exiting
through $\phi=\phi_{E}$ at any time is\begin{equation}
p_{\textrm{exit}}=\int_{0}^{\infty}dp_{\textrm{exit}}=-v_{*}(\phi_{E})\int_{0}^{\infty}P_{c}(\phi_{E},t)dt.\end{equation}
To compute $p_{\textrm{exit}}$ directly without knowledge of $P_{c}(\phi,t)$,
we define an auxiliary function\begin{equation}
Q(\phi)\equiv\int_{0}^{\infty}dt\, P_{c}(\phi,t),\end{equation}
so that \begin{equation}
p_{\textrm{exit}}=-v_{*}(\phi_{E})Q(\phi_{E}).\end{equation}
 It is easy to see that the function $Q(\phi)$ is a solution of\begin{equation}
\hat{L}_{c}Q(\phi)=P_{c}(\phi,t=\infty)-P_{c}(\phi,t=0)=-\delta(\phi-\phi_{0}),\label{eq:for Q}\end{equation}
since $P_{c}(\phi,t=\infty)=0$. (Note that the function $v_{*}(\phi)Q(\phi)$
satisfies a gauge-invariant equation, \begin{equation}
\partial_{\phi}\left[\partial_{\phi}\left(\frac{D}{v_{*}}(v_{*}Q)\right)-v_{*}Q\right]=-\delta(\phi-\phi_{0}),\end{equation}
which reflects the fact that $p_{\textrm{exit}}$ is a gauge-invariant
quantity.) The boundary conditions are the time-integrated conditions~(\ref{eq:bc J P}),\begin{equation}
\left.\partial_{\phi}\left(DQ\right)\right|_{\phi_{E}}=0,\quad Q(\phi_{\max})=0,\label{eq:Q bc}\end{equation}
where we have chosen the purely absorbing boundary condition at $\phi=\phi_{\max}$
because we are now interested in regions that never reach that boundary.
Rewriting Eq.~(\ref{eq:for Q}) as\begin{equation}
\partial_{\phi}\left[\partial_{\phi}\left(DQ\right)-v_{*}Q\right]=-\delta(\phi-\phi_{0}),\end{equation}
we may immediately integrate,\begin{equation}
\partial_{\phi}\left(DQ\right)-v_{*}Q=C_{1}-\theta(\phi-\phi_{0}),\end{equation}
where we note that $C_{1}=p_{\textrm{exit}}$ due to the boundary
condition at $\phi=\phi_{E}$. Then we obtain the general solution\begin{align}
Q(\phi) & =C_{2}\frac{1}{D(\phi)}\exp\left[\int_{\phi_{E}}^{\phi}\frac{v_{*}}{D}d\phi\right]\nonumber \\
+ & \frac{1}{D(\phi)}\int_{\phi_{E}}^{\phi}d\phi'\left(C_{1}-\theta(\phi'-\phi_{0})\right)\exp\left[\int_{\phi'}^{\phi}\frac{v_{*}}{D}d\phi\right].\end{align}
The constants of integration $C_{1,2}$ are determined from the boundary
conditions~(\ref{eq:Q bc}), and the final result is\begin{align}
p_{\textrm{exit}} & =\frac{-\frac{v_{*}(\phi_{E})}{D(\phi_{E})}\int_{\phi_{0}}^{\phi_{\max}}R(\phi')d\phi'}{R(\phi_{E})-\frac{v_{*}(\phi_{E})}{D(\phi_{E})}\int_{\phi_{E}}^{\phi_{\max}}R(\phi')d\phi'},\\
R(\phi) & \equiv\exp\left[\int_{\phi}^{\phi_{\max}}\frac{v_{*}}{D}d\phi\right].\end{align}

The above equations are valid for any $v_{*}(\phi)$ and $D(\phi)$;
we shall now specialize to the case of $k$-inflation. Since $v_{*}<0$
and is approximately constant, while $D(\phi)>0$ and grows with $\phi$,
the function $R(\phi)$ is approximately equal to 1 within a certain
(model-dependent) range of $\phi$, namely for $\phi_{R}<\phi<\phi_{\max}$,
where $\phi_{R}$ is determined by the condition\begin{equation}
1\sim\int_{\phi_{R}}^{\phi_{\max}}\frac{v_{*}}{D}d\phi\sim\frac{v_{0}\phi_{R}}{D(\phi_{R})}.\end{equation}
Since $R(\phi)$ quickly approaches zero for $\phi<\phi_{R}$, we
may neglect $R(\phi_{E})\ll1$ and express the exit probability as\begin{equation}
p_{\textrm{exit}}=C_{1}=\frac{\int_{\phi_{0}}^{\phi_{\max}}R(\phi)d\phi}{\int_{\phi_{E}}^{\phi_{\max}}R(\phi)d\phi}.\end{equation}
It follows that \begin{align}
p_{\textrm{exit}} & \approx1,\quad\phi_{0}<\phi_{R};\\
p_{\textrm{exit}} & \approx\frac{\phi_{\max}-\phi_{0}}{\phi_{\max}-\phi_{R}},\quad\phi_{0}>\phi_{R}.\end{align}
A worldline starting in the middle of the range $\left[\phi_{R},\phi_{\max}\right]$
will exit either at $\phi=\phi_{E}$ or at the upper boundary $\phi=\phi_{\max}$
with nearly equal probability. Therefore, we may interpret the range
$\left[\phi_{R},\phi_{\max}\right]$ as the {}``runaway diffusion''
regime. A typical comoving worldline will surely (with probability
$p_{\textrm{exit}}\approx1$) reach the reheating boundary $\phi_{E}$
only if the initial value $\phi_{0}$ is outside of the {}``runaway
diffusion'' regime, i.e.~$\phi_{0}<\phi_{R}$. The value $\phi_{R}$
is highly model-dependent, but generically $\phi_{R}\ll\phi_{\max}$.
Therefore, the existence of the {}``runaway diffusion'' regime limits
the independence of $k$-inflation models on initial conditions. 

Note that the {}``runaway diffusion'' regime is absent in models
of potential-driven inflation, where we have\begin{equation}
H(\phi)\approx\kappa\sqrt{V},\quad v_{*}(\phi)=-\frac{V'}{3H},\quad D(\phi)=\frac{H^{3}}{8\pi^{2}},\end{equation}
and the meaningful range of $\phi$ is $\phi<\phi_{\max}$, where
$\phi_{\max}$ is the Planck boundary determined by $V(\phi_{\max})\sim M_{Pl}^{4}$.
One finds\begin{equation}
R(\phi)=\exp\left[-\frac{8\pi^{2}}{3\kappa^{4}}\left(\frac{1}{V(\phi)}-\frac{1}{V(\phi_{\max})}\right)\right],\end{equation}
so $R(\phi)$ is negligibly small almost all the way until the boundary
$\phi=\phi_{\max}$. It follows that $\phi_{R}\sim\phi_{\max}$, so
the exit probability in potential-driven inflation is $p_{\textrm{exit}}\approx1$
for any $\phi_{0}$ within the allowed range $\phi_{0}<\phi_{\max}$.

\section*{Acknowledgments}

The authors are grateful to Florian Marquardt, Slava Mukhanov, Matthew
Parry, Ilya Shapiro, and Alex Vikman for stimulating and fruitful
discussions.

\appendix

\section{Attractor solutions in potential-driven inflation\label{sec:Attractor-solutions-in}}

In this appendix, we derive the attractor solution in models of potential-driven
inflation with a power-law potential $V(\phi)\propto\phi^{n}$. This
calculation serves as an illustration of the general method for analyzing
attractors that we developed in Sec.~\ref{sec:Attractor-solutions}.
While the existence of attractor behavior in potential-driven inflation
is long known, it has not been stressed that a unique attractor solution
can be singled out in the phase plane (see e.g.~Ref.~\cite{Liddle:1994dx}
where a statement is made to the contrary).

We consider a cosmological model of inflation driven by a minimally
coupled scalar field $\phi$ with a potential $V(\phi)$. The equations
of motion are\begin{align}
\ddot{\phi}+3H\dot{\phi}+V'(\phi) & =0,\\
\frac{1}{2}\dot{\phi}^{2}+V(\phi) & =\frac{3M_{Pl}^{2}}{8\pi}H^{2}\equiv\frac{1}{\kappa^{2}}H^{2}.\end{align}
Considering $\dot{\phi}$ as a function of $\phi$, i.e.~$\dot{\phi}=u(\phi)$,
we can reduce the above equations to\begin{equation}
\frac{du(\phi)}{d\phi}=-\frac{V'(\phi)}{u}-3\kappa\sqrt{\frac{u^{2}}{2}+V(\phi)}\equiv g(u,\phi).\end{equation}
For the purposes of further analysis, we assume that the potential
$V(\phi)$ has a monotonically growing behavior at $\phi\rightarrow\infty$,
going as $V(\phi)\propto\phi^{n}$ with $n\geq2$.

\subsection{Determining the attractor solution}

According to the criterion developed in Sec.~\ref{sub:Definition-of-attractor},
attractors are determined by their behavior at $\phi\rightarrow\infty$.
In the phase plane $(\phi,\dot{\phi})$, an attractor solution $\dot{\phi}=u_{*}(\phi)$
should have the property that neighbor solutions grow significantly
faster than $u_{*}(\phi)$ at $\phi\rightarrow\infty$. In the case
of potential-driven inflation with growing $V(\phi)$, we have (at
fixed $u$)\begin{equation}
g(u,\phi)\sim\max(V',\sqrt{V})\rightarrow\infty\;\textrm{as}\;\phi\rightarrow\infty.\end{equation}
Hence, every solution $u(\phi)$ grows at $\phi\rightarrow\infty$.
However, a generic trajectory grows exponentially fast for large $\phi$,
while we expect the attractor trajectory to grow slower than exponentially
(e.g.~polynomially). To make the attractor behavior apparent, we
change variables as $u(\phi)\equiv F(\phi)\tilde{u}(\phi)$, where
$F(\phi)$ is a fixed function that will be determined below. The
function $F(\phi)$ should have polynomial growth, $F(\phi)\sim\phi^{s}$,
such that the attractor solution is $\tilde{u}(\phi)\rightarrow\textrm{const}$
at $\phi\rightarrow\infty$. 

After the change of variables, the equation for $\tilde{u}(\phi)$
is\begin{align}
\frac{d\tilde{u}}{d\phi} & =\frac{g(F\tilde{u},\phi)-F'\tilde{u}}{F}\nonumber \\
 & =-\frac{V'}{F^{2}\tilde{u}}-3\kappa\sqrt{\frac{\tilde{u}^{2}}{2}+\frac{V}{F^{2}}}-\frac{F'}{F}\tilde{u}\equiv\tilde{g}(\tilde{u},\phi).\label{eq:u balance}\end{align}
 A solution $\tilde{u}(\phi)\rightarrow\textrm{const}$ at $\phi\rightarrow\infty$
will exist if $\tilde{g}(\tilde{u},\phi)$ has an {}``asymptotic
root'' $\tilde{u}_{0}<0$ such that\begin{equation}
\lim_{\phi\rightarrow\infty}\tilde{g}(\tilde{u}_{0},\phi)=0.\end{equation}
 Since $F'/F\sim\phi^{-1}$, the last term in Eq.~(\ref{eq:u balance})
is always dominated by the second term at fixed $\tilde{u}$ as $\phi\rightarrow\infty$.
Therefore, an asymptotic root will exist if the first two terms cancel
each other as $\phi\rightarrow\infty$. This requires that one of
the two sets of conditions hold in the asymptotic limit $\phi\rightarrow\infty$:\begin{eqnarray}
\frac{V'}{F^{2}}\sim\frac{\sqrt{V}}{F}, &  & \frac{\sqrt{V}}{F}\gg1,\\
 & \textrm{or}\nonumber \\
\frac{V'}{F^{2}}\sim1, &  & \frac{\sqrt{V}}{F}\ll1.\end{eqnarray}
For a power-law potential, $V(\phi)=\lambda\phi^{n}$, and $F(\phi)=\phi^{m}$,
it is straightforward to see that only the first set of conditions
can be met, which yields\begin{equation}
m=\frac{n}{2}-1,\quad F(\phi)=\phi^{\frac{n}{2}-1}\sim\frac{V'}{\sqrt{V}}.\end{equation}
 The {}``asymptotic root'' $\tilde{u}_{0}$ is\begin{equation}
\tilde{u}_{0}=-\frac{n\sqrt{\lambda}}{3\kappa}.\end{equation}
(The value of $\tilde{u}_{0}$ is negative since $\dot{\phi}<0$ for
large $\phi$.) With the choice $F(\phi)=\phi^{n/2-1}$, the attractor
is determined as the (unique) solution $\tilde{u}_{*}(\phi)$ which
approaches a constant as $\phi\rightarrow\infty$. Other solutions
grow exponentially with $\phi$.

In the original variables, the attractor solution is\begin{equation}
u_{*}(\phi)\approx F(\phi)\tilde{u}_{0}\sim\phi^{\frac{n}{2}-1}\;\textrm{as }\phi\rightarrow\infty.\end{equation}
 This is the well-known slow-roll attractor behavior in potential-driven
inflation, $\dot{\phi}\sim V'/\sqrt{V}$.

\subsection{Approximate expressions for the attractor}

A first approximation to the attractor solution is $\tilde{u}_{*}(\phi)\approx\tilde{u}_{0}=\textrm{const}$,
which corresponds (in the original variables) to the solution of the
equation\begin{equation}
u(\phi)=\frac{d\phi}{dt}=-\frac{V'}{3\kappa\sqrt{V}}.\end{equation}
This is the familiar slow-roll approximation. Higher-order asymptotic
approximations may be determined by iterating Eq.~(\ref{eq:u balance}),
considered as an equation for $\tilde{u}(\phi)$ with a given $d\tilde{u}/d\phi$.
After some algebra, the next-order approximation to the attractor
solution is found as\begin{equation}
u_{*}(\phi)\approx-\frac{V'}{3\kappa\sqrt{V}}\left[1+\frac{4VV^{\prime\prime}-3V^{\prime2}}{36\kappa^{2}V^{2}}\right].\end{equation}
As expected, this expression reproduces the slow-roll expansion of
Ref.~\cite{Liddle:1994dx} up to terms of first order.

\bibliographystyle{apsrev}
\bibliography{1}

\end{document}